\documentclass[manuscript, screen, nonacm]{acmart}
\AtBeginMaketitle{\authorsaddresses{}}
\AtBeginDocument{%
  }

\usepackage{tabularx}
\usepackage{float}
\usepackage{xcolor}
\usepackage{todonotes}
\usepackage{xcolor}
\usepackage{fontawesome}
\usepackage[most]{tcolorbox}
\usepackage{xspace}

\normalmarginpar

\newcommand{\reviewplan}[3]{%
  \noindent
  \todo[
    backgroundcolor=white,
    bordercolor=white,
    linecolor=orange!65!black,
    textcolor=black!70,
    size=\footnotesize,
    nolist
  ]{%
    \textcolor{orange!80!black}{\textbf{#1}}\\
    #2%
  }%
  {\small\textcolor{blue!70!black}{%
    \textbf{REVISION PLAN:} #3%
  }}%
  \space
}

\renewcommand{\reviewplan}[3]{}
\definecolor{boxcolor}{RGB}{238, 223, 204} %

\begin{document}

\title{Public Perceptions and Acceptance of Socially Intelligent AI Agents}

\author{Leena Mathur}
\email{lmathur@cs.cmu.edu}
\affiliation{%
  \institution{Carnegie Mellon University}
  \city{Pittsburgh}
  \state{Pennsylvania}
  \country{USA}
}

\author{Jenny T. Liang}
\affiliation{%
   \institution{Carnegie Mellon University}
  \city{Pittsburgh}
  \state{Pennsylvania}
  \country{USA}
}

\author{Vasudha Varadarajan}
\affiliation{%
   \institution{Carnegie Mellon University}
  \city{Pittsburgh}
  \state{Pennsylvania}
  \country{USA}
}

\author{Jimin Mun}
\affiliation{%
 \institution{Carnegie Mellon University}
  \city{Pittsburgh}
  \state{Pennsylvania}
  \country{USA}
}

\author{Xuhui Zhou}
\affiliation{%
  \institution{Carnegie Mellon University}
  \city{Pittsburgh}
  \state{Pennsylvania}
  \country{USA}
}

\author{Jana Schaich Borg}
\affiliation{%
  \institution{Duke University}
  \city{Durham}
  \state{North Carolina}
  \country{USA}
}

\author{Yonatan Bisk}
\affiliation{%
  \institution{Carnegie Mellon University}
  \city{Pittsburgh}
  \state{Pennsylvania}
  \country{USA}
}

\author{Louis-Philippe Morency}
\affiliation{%
 \institution{Carnegie Mellon University}
  \city{Pittsburgh}
  \state{Pennsylvania}
  \country{USA}
}

\author{Maarten Sap}
\affiliation{%
  \institution{Carnegie Mellon University}
  \city{Pittsburgh}
  \state{Pennsylvania}
  \country{USA}
}

\renewcommand{\shortauthors}{}

\begin{abstract}
AI researchers have been advancing socially intelligent AI agents (Social-AI) across embodiments, from chatbots to physical robots. As Social-AI agents move from labs into everyday settings, understanding public perceptions and acceptance of Social-AI agents becomes increasingly important to advance agents that support human well-being. We present findings from a mixed-methods survey of United States adults (N=200) that examined whether participants perceive current AI agents as socially intelligent, abilities signaling social intelligence in agents, layperson acceptance of agents across contexts and embodiments, and concerns about Social-AI agents. We find that participants attributed social intelligence to AI agents with judgments grounded in observable agent behavior. We identified a support-adoption gap in acceptability judgments, with participants supporting the existence of Social-AI agents for others far more than for their own use. We contribute a novel analysis of layperson concerns and reservations about Social-AI to inform responsible design, deployment, and evaluation of these technologies.
\end{abstract}




\maketitle

\section{Introduction}
Research in artificial intelligence has advanced the development and deployment of \textit{socially intelligent} AI (Social-AI) agents across embodiments.  Conversational chatbot agents are now used by millions of people every day  \cite{NBERw34255,handa2026enablingindependentresearch}, and physical robots have begun to interact more frequently with humans to provide assistance in homes \cite{ranganeni2024robots}, hospitals \cite{thomaz2023robots}, and other settings. Despite having different embodiments, both chatbots and robots in these situations must  ``read the room'' by accurately inferring human intent and responding appropriately within each unique social context. For example, a chatbot or robot that interrupts a human to offer assistance at an awkward time can be intrusive and ineffective \cite{10.1145/3290605.3300233}. As AI agents increasingly move from labs into real-world deployments in healthcare \cite{8226850,sharma2023human, kian2025engagement, spitale2025past}, education \cite{gordon2016affective, park2019model}, hospitality \cite{nakanishi2020continuous}, and service \cite{brown2024trash}, this ability to ``read the room'' and exhibit social intelligence has become increasingly important for the future of human-AI interactions that support human health and well-being. Towards this end, 
Social-AI research aims to build computational foundations for AI agents to 
sense, perceive, reason about, learn from, and respond to affective, behavioral, and cognitive dimensions of human  interactions \cite{mathur-etal-2024-advancing, lee2024towards}. However, progress towards technical competencies in Social-AI, such as social perception and reasoning, does not necessarily correspond with whether users will perceive these AI agents to be socially intelligent. An agent may demonstrate competencies that, in theory, relate to social intelligence (e.g., recognizing user frustration) without users broadly perceiving the agent to be \textit{socially intelligent}; conversely, users may attribute social intelligence to agents whose underlying technical competencies remain limited or brittle.

Prior human-computer interaction (HCI) research has shown that humans treat computers as social actors \cite{nass1994computers, gambino2020building}, form holistic impressions of artificial agents \cite{nass2000machines,paepcke2010judging, gambino2020building}, and anthropomorphize agents  \cite{abercrombie2023mirages}. These perspectives establish that people can perceive and respond socially to machines, but do not directly provide insight into how users perceive social intelligence in AI agents. For example, a user may say ``please'' to a chatbot they perceive socially without judging the chatbot to be socially intelligent if it repeatedly misinterprets sarcasm or other social signals. Understanding how users perceive social intelligence and their willingness to use Social-AI agents will inform and improve the design, deployment, and evaluation of Social-AI as an interactive, human-centered technology. Motivated by the need for this understanding of public perceptions of Social-AI, we contribute an original, empirical study of  layperson perceptions of social intelligence in current AI agents and AI agent abilities that laypeople associate with social intelligence. 

While Social-AI agents are often \textit{intended for} layperson users, decisions regarding the design and deployment of these user-facing systems are largely \textit{made by} a small number of researchers and developers. The HCI community has long emphasized the importance of studying interactive technologies from the perspective of users and within real-world deployment contexts \cite{10.7551/mitpress/4398.001.0001, takayama2008beyond, andalibi2025public}. As Social-AI agents move into everyday settings, decisions regarding \textit{where}, \textit{how}, or \textit{to whom} these agents should be deployed should depend not only on agent technical abilities, but also on how laypeople perceive these agents \cite{selbst2019fairness}. For example, a Social-AI agent that is welcomed in one context (e.g., hospital robot driving up to assist visitors) may be intrusive or inappropriate in another (e.g., a home robot asking a question during a family argument). Moreover, the ability for Social-AI agents to ``read the room'' and exhibit social intelligence has  information-access and privacy implications for users \cite{10.1145/642611.642635}, as agents may collect sensitive interaction data.  Responsibly advancing Social-AI requires understanding not only whether people accept these agents in deployments, but also how designers might characterize and address
the space of concerns and risks that laypeople perceive. Motivated by the need for this understanding of public acceptance of these agents, we additionally study layperson acceptance of Social-AI agents across  deployment scenarios and layperson concerns or reservations about Social-AI. 

In this research, we designed and conducted a mixed-methods survey of adults in the United States (N=200) to address four user-facing questions that explored layperson perceptions of social intelligence in both chat-based and embodied (robotic) agents: 

\noindent\hangindent=2.6em\hangafter=1
\textbf{RQ1}\hspace{0.5em}To what extent do laypeople believe that current AI agents have social intelligence?

\noindent\hangindent=2.6em\hangafter=1
\textbf{RQ2}\hspace{0.5em}What abilities exhibited by an AI agent might convince laypeople that the agent has social intelligence?

\noindent\hangindent=2.6em\hangafter=1
\textbf{RQ3}\hspace{0.5em}How is layperson acceptance of a Social-AI agent influenced by deployment context and agent embodiment? 

\noindent\hangindent=2.6em\hangafter=1
\textbf{RQ4}\hspace{0.5em}What concerns or reservations might laypeople have regarding risks or harms of Social-AI agents?

We draw on perspectives from \textit{Participatory AI}  and human-centered design, which argue that people affected by AI systems should be empowered with opportunities to inform how those systems are designed, developed, and deployed \cite{takayama2008beyond,birhane2022power, bondi2021envisioning, delgado2023participatory}. Our findings reveal that participants readily attributed the construct of social intelligence to AI agents, with these judgments primarily grounded in observable agent behaviors, rather than belief in these systems having internal agency or intent. Participants’ willingness to seek out a Social-AI service for themselves, actively avoid it, and be glad that it existed for others uncovers a \textit{support-adoption gap} with participants supporting the existence of Social-AI agents for others far more than for their own personal use.   The findings from our original, empirical study contribute an understanding of layperson perceptions and acceptance of socially intelligent AI that yield actionable considerations for the responsible design, development, and deployment of these technologies.

\section{Relevant Work}
\label{sec:related_work}

\subsection{Socially Intelligent AI Agents}
\label{subsec:si_ai}

Social intelligence as a distinct form of intelligence has been studied for over a century in cognitive science and the social sciences \cite{thorndike1937evaluation, Kihlstrom_Cantor_2000}. Though there is debate in this ongoing research area, frameworks for social intelligence commonly view this construct as involving the following competencies: social perception, social knowledge and memory, social reasoning, theory-of-mind, and social interaction \cite{turner1988theory,weis2005social, conzelmann2013new}. Social-AI research has drawn upon these frameworks to build computational foundations for agents with social intelligence \cite{gweon2023socially,mathur-etal-2024-advancing, lee2024towards, li2024social}. Research towards this goal has spanned multiple computing communities and resulted in evaluation benchmarks,  environments,  algorithms, and design principles to improve social intelligence competencies in AI systems \cite{sap2019social,zhou2024sotopia,mathur-etal-2025-social, 10.1145/3772318.3790936,wang2025rethinking}. 

However, AI agents can satisfy a given benchmark without users recognizing the agent as socially intelligent. In turn, users may attribute social competencies to AI agents that do not exhibit strong performance on scientific benchmarks for social intelligence \cite{weizenbaum1983eliza, natale2019if}. Technical abilities of Social-AI agents and user perception of these abilities are both important in deployment and governance decisions. Our work studies social intelligence as a \textit{perceived} construct in current AI agents by examining layperson perceptions of agents' abilities. Prior work has studied perceived social intelligence in robots \cite{barchard2020measuring} and elicited expert perspectives on characteristics that socially intelligent conversational agents should exhibit \cite{brinkschulte2022perspectives}. Our study complements these perspectives by contributing empirical knowledge about the behaviors and abilities laypeople associate with social intelligence in current AI agents.

\subsection{Participatory AI, Responsible AI, and Public Acceptance of AI}
\label{subsec:participatory_ai}
A growing body of research in Participatory AI has contended that people affected by AI should be able to inform the development and deployment of these emerging technologies \cite{bondi2021envisioning,birhane2022power,delgado2023participatory}. Participatory AI is especially relevant for Social-AI agents that are designed to interact with layperson users in everyday, often sensitive, social interaction contexts. Recent studies have designed surveys to elicit layperson perceptions of AI capabilities, risks, and acceptable use cases and document gaps between expert and public perceptions of these systems  \cite{mun2024particip,ullstein2025participatory,mushkani2025co,mun2025not, brauner2026charting}. 

Parallel lines of work in human-machine interaction have studied user acceptance of social agents, ranging from public attitudes towards acceptable robot roles \cite{takayama2008beyond} to studies of how particular user groups respond to social agents, such as elderly user perceptions of assistive social robots
\cite{heerink2010assessing}. More recent works have found that public perceptions of emotionally intelligent AI vary across deployment contexts, user, and data-related dimensions \cite{andalibi2025public, ingber2025distinguishing}. Our survey complements these prior works by examining social intelligence as a holistic perceived construct in AI agents, 
contributing new findings regarding how laypeople envision social intelligence in AI, abilities that make an agent appear socially intelligent, and layperson acceptance and concerns. Our study contributes a public-facing account of Social-AI perceptions, bridging technical perspectives on Social-AI with Participatory AI frameworks for responsible advancement of this technology. 

\section{Study Methodology}
\label{sec:methods}

To examine how layperson participants perceive and reason about social intelligence in AI agents, we conducted a mixed-methods survey of adults in the United States (N=200). Participants were recruited through Prolific and the survey was administered through Qualtrics in April 2026. Informed by prior surveys designed to assess public perceptions of AI \cite{mun2024particip,bondi2021envisioning}, we recruited a sample of laypeople with eligibility restricted to adults residing in the United States and fluent in English. Participants were compensated at \$12 per hour (pro-rated), the mean survey completion time was 19.53 minutes, and the median survey completion time was 17.5 minutes. All study procedures were approved by an IRB. 

\begin{figure}[t]
    \centering
\includegraphics[width=\linewidth]{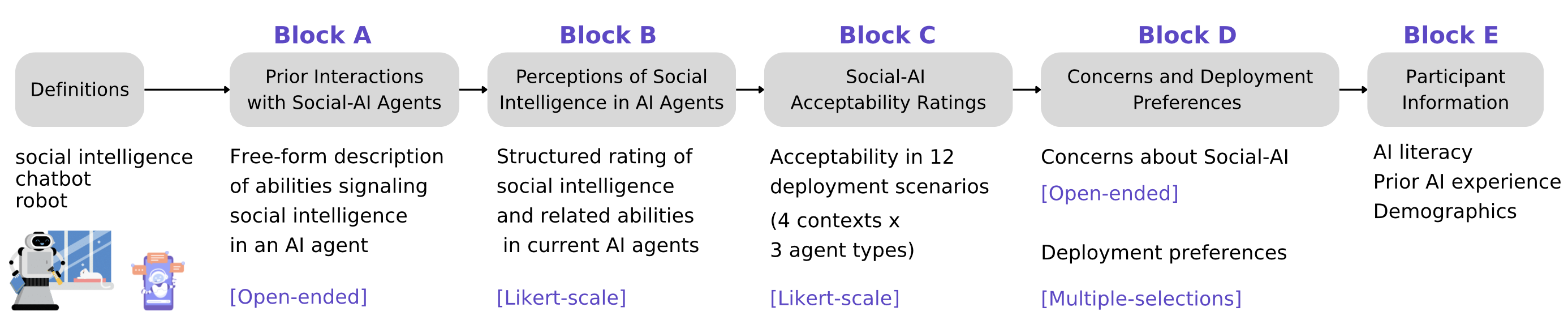}
    \caption{Overview of the survey instrument. After reading definitions of key terms, participants  completed five sequential blocks.}
    \label{fig:flow}
\end{figure}

\subsection{Survey Instrument}
\label{subsec:survey}
The survey instrument was designed with five question blocks presented in a fixed order. Block A elicited participants' prior experience and open-ended interpretations of social intelligence in AI agents, and Block B collected participants' structured ratings of current AI agent capabilities (addressing RQ1 and RQ2). Block C surveyed participants' judgments about Social-AI acceptability across deployment contexts and agent types (addressing RQ3). Block D surveyed participants' concerns, reservations, and preferences about Social-AI (addressing RQ4). Block E, the final block, collected information about participants. The survey design was intended to minimize ambiguity for non-expert respondents by grouping related questions into blocks and moving from open-ended interpretations to structured ratings \cite{marsden2010handbook, krosnick2017questionnaire}. 
Within each block, the Likert-scale items and scenario-based questions were presented in a randomized order to reduce order effects. The  survey flow is visualized in Figure \ref{fig:flow}.

Before taking the survey, participants read short definitions of the terms \textit{AI agent} and \textit{social intelligence} and viewed visual examples of chatbots and robots (provided in Appendix Figure \ref{fig:anchors}). Following prior HCI survey design \cite{andalibi2025public}, 
we provided these definitions and examples at the beginning of the survey to establish shared reference points for non-expert participants who may be unfamiliar with these terms, before they reasoned about Social-AI. These definitions preceded Block A; therefore, we note that participants' responses to the survey should be interpreted relative to the study's scope and framing, rather than as entirely unprompted conceptions of Social-AI. From the widely used MUFaSAA dataset of social robot embodiments \cite{dennler2023design}, we selected robot samples that were appropriate for the scenario contexts in Block C. 
Our robot samples are images of real-world mobile robots, are plausibly able to approach a person, do not appear toy-like, span both humanoid and non-humanoid embodiments, and have been highly rated by users in perceived competence \cite{dennler2023design}. These chatbot and robot images serve as anchors to help participants reason about survey questions from a shared visual understanding of agent embodiments.

\paragraph{\textbf{Block A: Prior Interactions with Social-AI Agents}} The first question block had participants indicate whether or not they had interacted with an AI agent they perceived as socially intelligent. Participants who answered ``yes" described the agent, identified its embodiment from multiple choice options (chatbot with no physical body, humanoid physical robot, non-humanoid physical robot), and answered an open-ended question about abilities, behaviors, and characteristics that made the agent appear socially intelligent. Participants who answered ``no" answered a parallel open-ended question about abilities, behaviors, and characteristics that an AI agent would need to convince them that it had social intelligence. We placed this open-ended question before the structured Likert-scale agent capability ratings (Block B) so that participants could first describe their own interpretations of social intelligence before seeing the structured dimensions selected by researchers \cite{krosnick2017questionnaire, marsden2010handbook}.

\paragraph{\textbf{Block B: Perceptions of Social Intelligence in AI Agents}} The second question block measured participants' perceptions of social intelligence in current AI agents. Participants were asked to consider the most advanced AI agents they were aware of, even if they had not personally interacted with them, and rate eight statements on a 5-point Likert scale (1 = strongly disagree, 5 = strongly agree). One item elicited a holistic judgment of perceived social intelligence (``Current AI agents have social intelligence.'') and the remaining items assessed participant perceptions of agent capabilities that are typically associated with Social-AI by researchers \cite{barchard2020measuring, brinkschulte2022perspectives, mathur-etal-2024-advancing}: the extent to which current AI agents are pleasant and sociable to interact with, recognize and respond appropriately to human emotion, 
take initiative during conversations and offer useful assistance, 
build rapport and trust with humans, feel natural and human-like, understand what is ethically right and wrong, and act with purpose and follow their own intentions. We analyze the holistic social intelligence item directly and examine the extent to which the seven related agent capability ratings are associated with that holistic judgment. 

\begin{table*}[t]
\centering
\small
\setlength{\tabcolsep}{3pt}
\renewcommand{\arraystretch}{1.08}
\begin{tabular}{p{0.09\linewidth}p{0.10\linewidth}p{0.12\linewidth}p{0.63\linewidth}}
\hline
\textbf{Setting} & \textbf{Stakes} & \textbf{Agent role} & \textbf{Scenario context stem} \\
\hline
Home & Lower & Work Support & An <agent> notices when you are overwhelmed with work and suggests a break. \\
Home & Higher & Fall Alert & An <agent> alerts you when an elderly parent falls and appears to need assistance. \\
Hospital & Lower & Wayfinding & An <agent> notices when you appear confused and provides directions. \\
Hospital & Higher & Distress Alert & An <agent> monitors your emotional state and alerts your care team about  distress. \\
\hline
\end{tabular}
\caption{The 4 scenario contexts used to construct the 12 scenario conditions surveyed for acceptability. Each scenario context combined a setting, relative stakes, and agent role. Agent roles included Work Support, Fall Alert, Wayfinding, and Distress Alert. Each scenario context was crossed with 3 agent types: chatbot, autonomous robot, and teleoperated robot. Exact participant-facing wording for all scenario conditions appears in Appendix Table \ref{tab:scenario_wording_appendix}.}
\label{tab:scenario-use-cases}
\end{table*}

\paragraph{\textbf{Block C: Social-AI Acceptability Ratings}} 
The third question block assessed participants' responses to Social-AI agents across deployment situations. Participants read and rated 12 scenario conditions. Each scenario condition paired one of four \textit{scenario contexts} with one of three \textit{agent types}. The four scenario contexts featured concrete, deployment situations that differed in  \textit{setting} (home and hospital), \textit{relative stakes} (lower stakes and higher stakes), and an agent \textit{role} associated with each setting-by-stakes combination. Home and health-related contexts are also, currently, among the most common real-world Social-AI deployment domains \cite{hurst2020social,thomaz2023robots,henschel2021makes,spitale2025past}, motivating our choice to scope our scenario contexts to home and hospital settings. In addition, home and hospital settings differ in privacy expectations and normative behaviors.
Our scenario contexts represent a set of plausible Social-AI deployment situations.   

These four scenario contexts were crossed with three agent types: chatbot, autonomous robot, and teleoperated robot.
We minimally adapted the wording for each agent type to ensure plausibility. We systematically varied agent type, since whether a social agent is disembodied (e.g., chatbot) or embodied (e.g., physical robot) can influence how people respond to and reason about the system \cite{deng2019embodiment}. Within our physical robot agent types, we further distinguished between \textit{autonomous robots} and \textit{teleoperated robots} (a physical robot operated remotely by a person). We included teleoperated robots because 
users may perceive social presence and concerns differently when a robot is being remotely controlled by a person \cite{kaminski2016averting,parts2025systematic}. The scenario contexts are summarized in Table \ref{tab:scenario-use-cases}, and wording for all 12 scenario conditions is in Appendix Table \ref{tab:scenario_wording_appendix}. We study participant responses at the level of the complete scenario context; disentangling the contribution of individual scenario dimensions is outside the scope of the study. 

For each scenario condition, participants rated three statements: whether they would actively seek out the service for themselves, actively avoid the service for themselves, and be glad the service existed for others who wanted to use it. Because participants were instructed to evaluate hypothetical use cases, we note that low reported willingness to seek out a service for oneself does not necessarily indicate rejection of the technology; it may reflect limited personal need. The avoidance item more directly captures a participant's hypothetical rejection of the Social-AI agent. Participants were instructed to assume that the agent could perform at least as well as an average human, communicate through voice/audio, and store and process interaction data.  The instructions, definitions, and visual examples shown to participants before the scenario ratings are provided in Appendix Figure \ref{fig:block_4}.
This assumption of agent competence was intended to hold expected task performance constant across hypothetical scenarios, so that participants could focus on evaluating the described deployment, rather than considering whether an agent would be capable of performing the task in the first place. We did not include a non-Social-AI control condition as our goal was not to study whether Social-AI agents are preferred over alternative technologies. Scenarios were presented in randomized order and included a diagnostic attention-check item instructing participants to select Neutral for all three statements.

\paragraph{\textbf{Block D: Concerns and Deployment Preferences}} The fourth question block surveyed participant concerns and reservations about Social-AI. Participants first answered an open-ended question asking them to describe concerns or reservations about the risks or harms of Social-AI or to explicitly write a statement declaring they had no concerns.  After completing the open-ended response, participants answered a set of multiple-response questions about data usage and storage preferences for  Social-AI robots in the home. In the questions, we varied whether the robot was autonomous or teleoperated and whether interaction data would be stored locally or in the cloud. For the teleoperated conditions, participants were explicitly told in the definition of teleoperated that interaction data would be visible to the human teleoperator. In each case, participants could select continuous speech, speech only during active interaction, continuous video, video only during active interaction, or none of the above. These questions were designed to study participant preferences and boundaries around sensing modalities, sensing durations, data storage location, and human access to interaction data. 

\paragraph{\textbf{Block E: Participant Information}} The final question block surveyed participants' prior AI experience, AI literacy, and demographic characteristics. Participants reported how frequently they interacted with chatbots and robots in their lives and, if applicable, the types of agents they regularly used. Participants completed a 7-item \textit{AI Literacy Scale}, listed in Table \ref{tab:ai_literacy_scale}, adapted from prior works \cite{mun2024particip, wang2023measuring} that measured self-reported ability to identify AI technologies, use AI tools, choose appropriate AI tools for tasks, use AI ethically, attend to privacy and security risks, recognize potential abuse of AI technology, and understand limitations of AI. These items measure self-assessments and reported practices, rather than demonstrated knowledge. The Qualtrics survey collected participants' education, occupation, and race. Education and occupation items allowed participants to enter ``N/A" if they preferred not to disclose; the race item was optional. We supplemented surveyed  data with Prolific-provided age, sex, and country-of-residence. 

\subsection{Participants}
\label{subsec:participants}
Our final sample included 200 participants and ranged from 18 to 78 years old (M=40.9, SD=12.8). Prolific's data recorded the sample to be approximately gender-balanced (53\% men, 45\% women, 2\% did not  disclose). Our sample size was selected to support descriptive estimates and within-participant comparisons across the 12 scenario conditions, as well as  thematic analysis of participant responses. 
The within-participant scenario design yields 2400 acceptability ratings (200 participants x 12 scenarios) for the mixed-effects analysis. A sensitivity calculation for a single paired t-test indicated a detectable standardized within-participant mean difference of $d_z \approx 0.20$ (80\% power, two-sided $\alpha = .05$). The observed 95\% confidence intervals for condition means had margins of $\pm 0.03$--$0.05$ on the normalized 0-1 scale. Given our sample size, we note that low-prevalence qualitative codes should be viewed as exploratory.

Most participants reported having prior experience with chatbots: 50\% of participants reported using a chatbot daily, and 29\% reported using a chatbot a few times a week. Prior experience with physical robots was less common among participants, with 72.5\% of participants never having used a robot. Accordingly, we note that participant judgments of the robot scenarios in Block C should be interpreted as their perception of hypothetical interactions with robots, rather than judgments grounded in their own prior experience. Participants represented varied professional backgrounds. We mapped responses to  Bureau of Labor Statistics Standard Occupational Classification (SOC) major groups using keyword rules \cite{bls_soc_2018}. Occupational responses included professions spanning business, management, sales, education, healthcare, transportation, and service roles, in addition to students and people not currently employed. This occupational diversity supports interpreting the sample as a layperson sample. Following recent surveys of public perceptions of AI \cite{mun2024particip}, we refer to this sample as laypeople since participants were recruited without an AI-expertise requirement. While we obtain original insights and empirical knowledge from our N=200 sample, we do not use the sample to estimate population-level public opinion. 

\subsection{Data Quality Validation}
\label{subsec:quality}
Participants were instructed to answer questions in their own words and not use ChatGPT or other writing assistants. We enabled Prolific's automated LLM-use detection on free-text responses from Qualtrics, which flagged submissions exhibiting patterns that signaled potential LLM-generated text. Two participants selected the instructed response for two of the three attention-check statements, but selected a different response for the remaining statement. Their responses were otherwise coherent, complete, and not flagged as AI-written by Prolific's detector.  Because the attention check was used diagnostically, rather than as an automatic exclusion criterion, we chose to retain these participants after authors independently validated the response quality. In Block B, one participant omitted ``Are pleasant and sociable to interact with'' and ``Act with purpose and follow their own intentions,'' resulting in N=199 for these two items and N=200 for the remaining six items. All participants supplied the seven AI literacy ratings (N=200). Quantitative analyses of the Block B Likert-scale ratings used complete available responses across all eight items (N=199). Qualitative analyses of Block A and Block D open-ended responses use statements from all participants (N=200). Authors manually validated all open-ended responses during qualitative coding. 

\subsection{Measures and Internal Consistency}
\label{subsec:scale}
From Block B, we analyzed the single holistic social intelligence rating and the seven related capability ratings separately. For the seven-item self-reported AI literacy measure, we normalized each item to 0-1 and averaged the seven ratings per participant to obtain a composite AI literacy score for each participant. We validated the internal consistency of the AI literacy scale with Cronbach's $\alpha$ and McDonald's $\omega$ \cite{cronbach1951coefficient, mcdonald2013test, dunn2014alpha}. Both demonstrated acceptable-to-high internal consistency ($\alpha$ = .801, $\omega$ = .804, N=200). These internal consistency estimates support combining the seven AI literacy items into a composite self-reported AI literacy score for our analyses.

\subsection{Quantitative Analysis}
\label{subsec:quant}
We analyzed Likert-scale and closed-ended survey responses with descriptive statistics, response percentages, and within-participant comparisons. For Block B Likert scale ratings, we computed two-sided Spearman correlations \citep{spearman1961proof} between the holistic social intelligence rating and each of the seven related capability ratings, applying Benjamini--Hochberg correction across these seven comparisons \cite{benjamini1995controlling}.

For scenario acceptability ratings (Block C), we normalized each 1--5 Likert rating to a 0--1 scale as $(x-1)/4$. We analyzed Seek-for-self, Avoid-for-self, and Glad-for-others separately. Avoidance was not reverse-coded; we treat higher ratings as representing greater active avoidance of the Social-AI agent in each given scenario. For descriptive condition means, we calculated 95\% confidence intervals as the mean $\pm1.96$ standard errors across participants. We used linear mixed-effects models to account for repeated scenario ratings. For each outcome, a model included four-level scenario context, three-level agent type, their interaction, and a participant random intercept. We report omnibus Wald $\chi^2$ tests for scenario context, agent type, and their interaction; the scenario and agent-type tests average across levels of the other factor. The nine omnibus tests use unadjusted p-values. These models compare bundled deployment scenarios and our study was not scoped to  isolate independent effects of setting, stakes, or role. We also examined how AI literacy and perceived social intelligence related to scenario ratings and compared participants’ data sensing preferences reported in Block D. Statistical tests for these analyses are reported in the corresponding results and Appendix tables.
 
We defined the Support-Adoption gap as Glad-for-others minus Seek-for-self and summarized its overall mean descriptively. For each of the 12 scenario conditions, we compared the same participants' two ratings using a two-sided Wilcoxon signed-rank test \citep{wilcoxon1945individual}, omitting zero differences and applying Holm correction across the 12 comparisons.

\subsection{Qualitative Coding Methodology}
\label{subsec:grounded_theory}
We analyzed open-ended responses for RQ2 and RQ4 using a multi-label qualitative coding process \cite{corbin1990grounded}. Two authors independently reviewed an initial development set of 20 responses and created independent codebooks, inductively generating codes. The two authors then met to merge overlapping codes and clarify definitions. The codebooks were then independently applied by both authors to code a new validation set of 30 responses. Human-human agreement on the validation set was strong (RQ2: Krippendorff's $\alpha$ = 0.87, RQ4: Krippendorff's $\alpha$ = 0.89). 
Following recent studies for qualitative analysis of AI perception surveys \cite{mun2024particip}, we used codebook-constrained LLM-assisted coding to apply the finalized RQ2 and RQ4 codebooks to the remaining responses. We used Anthropic \texttt{claude-opus-4-7} in a single-run closed-coding setup. The prompts provided the codebook definitions and coding rules and instructed the model to return only JSON arrays of valid code names. Agreement across human coders and the LLM on the validation set was strong (RQ2: Krippendorff's $\alpha$ = 0.82, RQ4: Krippendorff's $\alpha$ = 0.84). We manually audited the final coded dataset for missing, invalid, or inconsistent labels. As each response could receive multiple codes, we note that reported code percentages do not sum to 100\%. 

\subsection{Limitations}
\label{sec:limitations}
Our study was designed to survey participants' perceptions of Social-AI with a sample that included English-speaking adults  in the United States, recruited through Prolific (N=200) in April 2026. The online survey platform and sample size limit generalizability regarding claims about national public opinion. Our survey captures perceptions at a single point in time, and public attitudes will likely evolve as AI systems become more widely deployed. Our findings should be interpreted as original evidence about how a diverse group of US adult laypersons reasoned about Social-AI. 

Participants were provided with definitions of AI agents and social intelligence, as well as images of chatbots and robots, before answering the survey. These definitions and images were needed to establish a shared reference point for non-expert respondents, but they may have shaped how participants interpreted the construct of social intelligence. Future work could examine how responses differ with alternative definitions.

To keep the survey completion time and cognitive load manageable for layperson participants, our scenario design paired scenario contexts (settings, stakes, agent roles) with agent types. This design choice treated scenario context as a bundled construct, resulted in 12 scenarios rated by each layperson participant, and allowed us to compare acceptability across concrete deployment situations. This design limits the ability to isolate independent effects within scenario contexts. We believe that the survey instrument and methodology contributed by our research will motivate and inform future work to investigate broader participant pools and Social-AI deployment scenarios.

\begin{figure*}
    \centering
    \includegraphics[width=\linewidth]{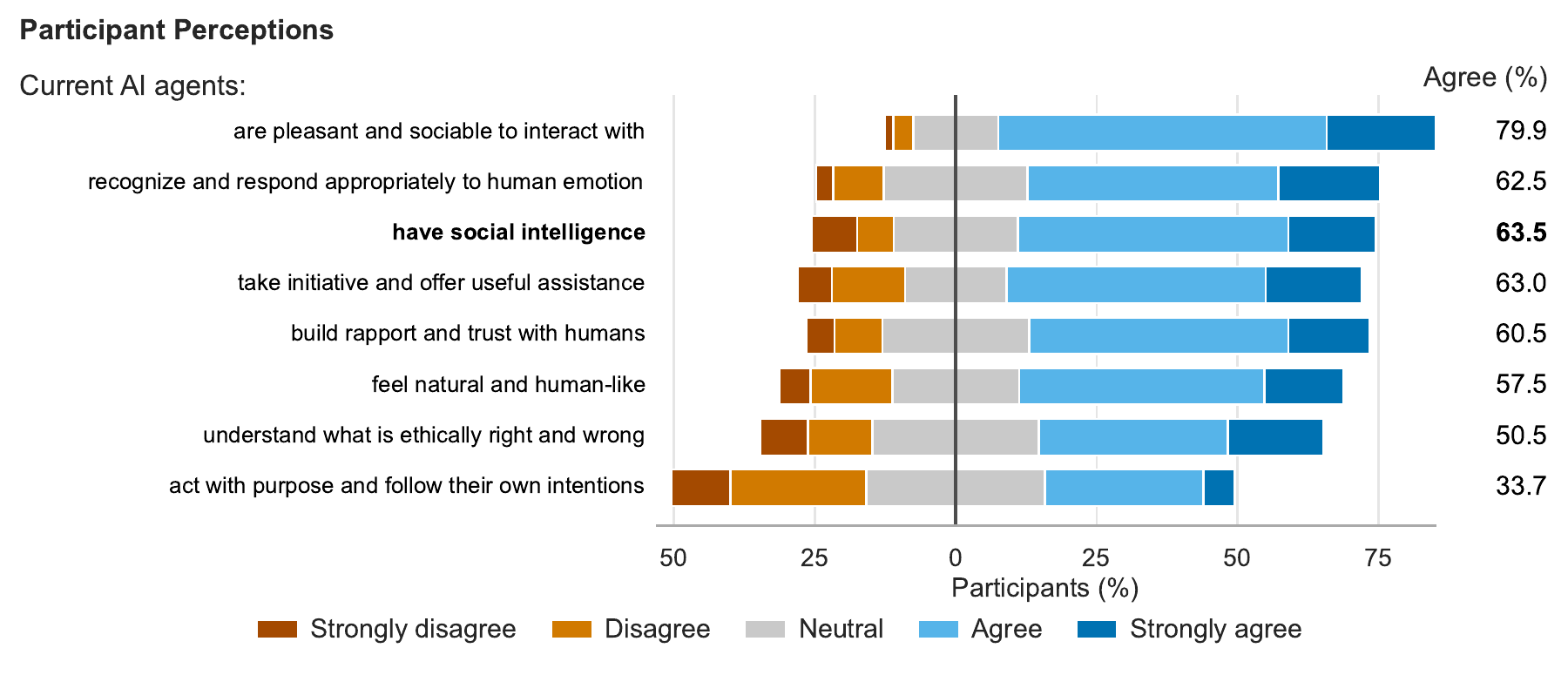}
    \caption{Distribution of participant agreement with statements about current AI agents from the Likert scale items in Block B. The holistic social intelligence item is bolded. Responses are shown as a diverging Likert plot centered at the neutral response category. The percentages of participants selecting Agree or Strongly Agree are on the right side of the figure.}
    \label{fig:rq1}
\end{figure*}

\section{Results and Discussion}
\label{sec:findings}

We align our discussion of findings with our research questions to  examine whether participants perceive current AI agents as socially intelligent (RQ1),   abilities signaling social intelligence in AI agents (RQ2), layperson acceptance of agents across contexts and embodiments (RQ3), and participants' concerns about Social-AI agents (RQ4). 

\subsection{Perceived Social Intelligence of AI Agents}
\label{subsec:rq1}
The majority of participants reported having interacted with an AI agent they perceived to be socially intelligent (89\%). Participant perceptions were primarily anchored in prior experience with chatbots; among participants who reported interacting with agents they perceived as socially intelligent, 97.2\% identified the agent as a \textit{chatbot} and 2.8\% identified the agent  as a \textit{physical robot}. This distribution across embodiments aligns with the current availability of AI systems; chatbots are widely accessible to millions of people within the general public, while physical robots are less commonly encountered in daily life due to higher barriers to entry and deployment costs \cite{blut2021understanding}. Findings in this section should be interpreted in light of participants' greater familiarity with chatbots than with physical robots.  

We analyzed participants' responses to the eight-item 
Likert scale in Block B. In Figure 
\ref{fig:rq1}, we visualize the distribution of participant agreement with statements about these eight items. In Appendix Table \ref{tab:rq1_item_summary}, we report the item-level summary statistics across the participant pool; M is their mean Likert rating for each item. Participants rated current AI agents highest on qualities such as being pleasant and sociable (M = 3.95), recognizing and responding to human emotion (M = 3.66), building rapport and trust with humans (M = 3.57), and taking initiative when offering useful assistance (M = 3.55). For the holistic statement about current AI agents having social intelligence, 63.5\% of participants selected \textit{Agree} or\textit{ Strongly Agree} (M = 3.57). Ratings were lowest for the ability of agents to act with purpose and intent (M = 2.94); responses to this item 
were split almost evenly across agreement, neutral, and disagreement. 

Participants' perceptions of agent abilities were positively correlated with their holistic rating of agent social intelligence. We computed item-level correlations between  ``Current AI agents have social intelligence'' and each of the other items (Appendix Table \ref{tab:rq1_item_correlations}). The largest observed correlation was with recognizing and responding appropriately to emotions ($\rho$ =.636), followed by natural and human-like behavior ($\rho$ =.560), pleasant and sociable interaction ($\rho$ =.555), understanding what is ethically right and wrong ($\rho$ =.549), and  building rapport and trust with humans ($\rho$ =.530). The weakest observed correlation was for acting with purpose and intent ($\rho$ =.367; all FDR-adjusted $p<.001$). 

\paragraph{\textbf{Implications of RQ1 Findings}} 
While advancing social intelligence in AI is often treated as a long-term goal for AI research from a scientific perspective \cite{fei2022searching, gweon2023socially, mathur-etal-2024-advancing},  \textit{we find that laypeople are already attributing the construct of social intelligence to the AI systems they have encountered}. However, participant ratings reveal an uneven attribution of agent capabilities towards this holistic perception of social intelligence. Layperson attributions of social intelligence to current AI agents focus primarily on observable interaction behaviors (e.g., recognizing and responding to human emotion, appearing pleasant and sociable), rather than assumptions of agents having intent or their own agency. Prior work warns that human-like language and sociability in AI agents can shape the mental capacities people attribute to these agents \cite{abercrombie2023mirages, chen2026presenting} and lead to risks such as overtrust and relational expectations that current AI cannot fulfill. These concerns depend on what users believe an agent understands, not solely on the technical capabilities it possesses. \textit{Our findings suggest that public-facing AI deployment and governance issues related to Social-AI are likely to arise before AI systems satisfy robust scientific definitions of social intelligence and must be proactively addressed.}

\subsection{Abilities Signaling Social Intelligence in AI Agents}
\label{subsec:rq2}
We analyze the open-ended responses from Block A to learn which abilities participants associate with social intelligence in AI agents. Participants who had encountered an AI agent they perceived as socially intelligent (89\%) described what made the agent seem socially intelligent, while those who had not (11\%) described what abilities an agent would need to convince them it had social intelligence. As both groups described the same underlying construct in these responses (agent abilities signaling social intelligence), we pooled responses for qualitative analysis to study the range of AI agent abilities mentioned by participants. We coded responses with an inductively developed multi-label codebook (Table \ref{tab:rq2_codebook}).

\begin{table}
\centering
\small
\setlength{\tabcolsep}{4pt}
\renewcommand{\arraystretch}{1.05}
\begin{tabular}{p{0.21\linewidth}p{0.21\linewidth}p{0.42\linewidth}rr}
\hline
\textbf{Ability Cluster} & \textbf{Code} & \textbf{Definition} & \textbf{Count} & \textbf{\%} \\
\hline
social actor skills & navigating conversations & Manages conversational flow with adaptive responses. & 72 & 36.0 \\
social actor skills & humanlike communication & Communicates in a natural, non-robotic manner. & 63 & 31.5 \\
social actor skills & social norms & Follows social conventions like politeness and interest. & 29 & 14.5 \\
\hline
prosocial behavior & expressing empathy & Communicates care, compassion, warmth toward the user. & 54 & 27.0 \\
prosocial behavior & helpful & Proactively offers unprompted help or suggestions. & 50 & 25.0 \\
prosocial behavior & building trust & Acts as a trustworthy conversational partner. & 6 & 3.0 \\
\hline
situational understanding & current context awareness & Infers user intent or situation in current exchange. & 60 & 30.0 \\
situational understanding & longitudinal awareness & Retains information across sessions over time. & 24 & 12.0 \\
\hline
affective capabilities & recognizing emotions & Detects user's emotional state or affective tone. & 59 & 29.5 \\
affective capabilities & expressing emotions & Conveys its own emotions or mirrors the user's. & 19 & 9.5 \\
\hline
functional capabilities & task competence & Performs domain-specific tasks accurately and usefully. & 47 & 23.5 \\
functional capabilities & logical structure & Produces well-structured, easily digestible responses. & 6 & 3.0 \\
\hline
n/a & rejecting the premise & Denies any AI agent could have social intelligence. & 2 & 1.0 \\
\hline
\end{tabular}
\caption{RQ2 codes for abilities signaling perceived social intelligence in AI agents. Representative quotations are provided in the Appendix Table \ref{tab:rq2_representative_quotes_appendix}. Responses could receive multiple codes, so percentages do not sum to 100\%.}
\label{tab:rq2_codebook}
\end{table}

\begin{figure}
    \centering
    \includegraphics[width=\linewidth]{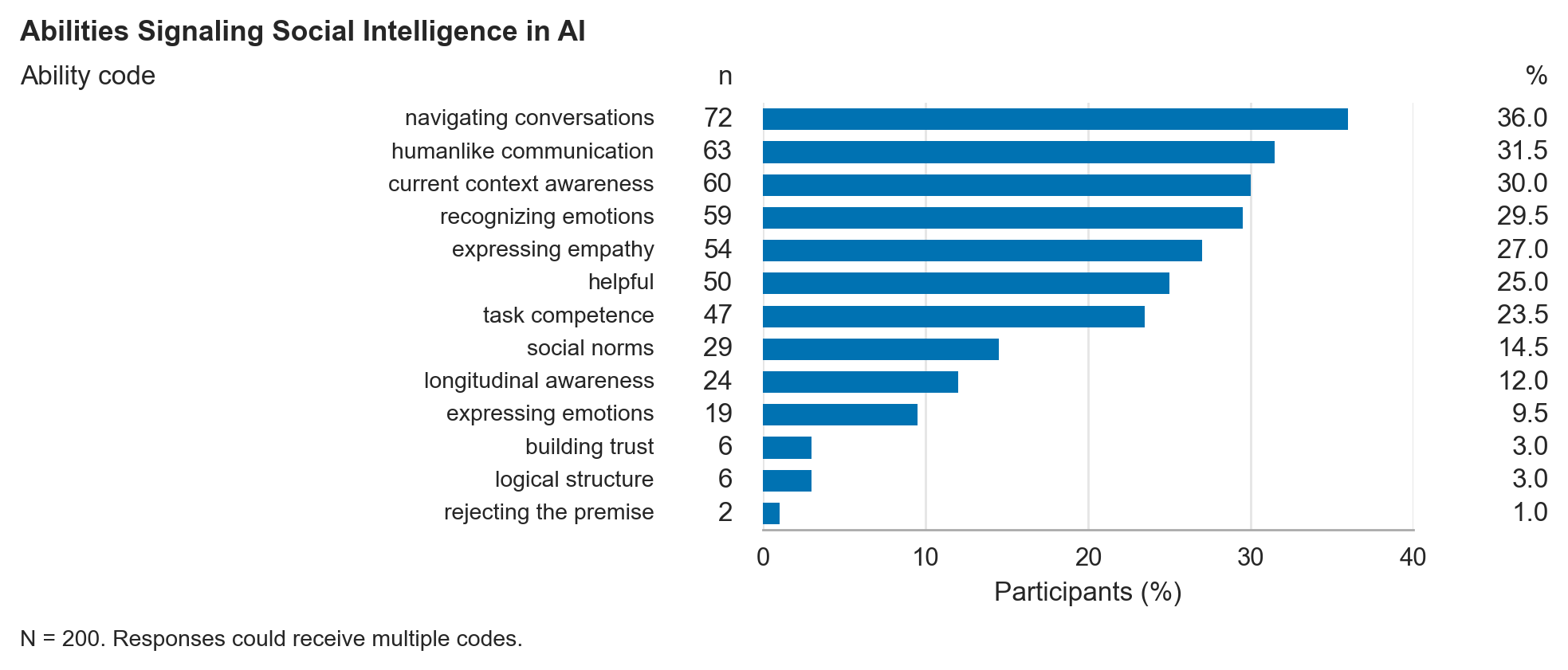}
    \caption{Distribution of RQ2 codes that participants identified as abilities signaling social intelligence in AI agents. Bars show code prevalence across open-ended responses. Responses could receive multiple codes.}
    \label{fig:rq2_codes}
\end{figure}

Participants associated social intelligence in AI agents with a broad set of agent abilities. The most frequently occurring codes were navigating conversations (36.0\%), humanlike communication (31.5\%), current context awareness (30.0\%), recognizing emotions (29.5\%), expressing empathy (27.0\%), proactively offering help (25.0\%), and exhibiting task competence (23.5\%). 
The less frequent, but recurring, codes related to following social norms (14.5\%), retaining information across sessions (12.0\%), and the agent expressing its own emotions (9.5\%). The distribution of RQ2 codes, in order of prevalence, is visualized in Figure \ref{fig:rq2_codes}. Overall, these codes span multiple categories in the codebook: social actor skills, situational understanding, affective capabilities, prosocial behavior, and functional capabilities.

These findings demonstrate that, collectively, participants did not equate social intelligence with a single ability; they, instead, reasoned about social intelligence as a multi-faceted construct. These qualitative findings align with our RQ1 finding regarding the divergence between perceived social intelligence in current AI agents and perceived agency in these agents.  
Most participants described social intelligence in behavioral terms regarding what the agent could \textit{do} in an interaction, as opposed to its internal state. Two participants rejected the premise that any AI agent could be socially intelligent and  centered this perspective on the view that social intelligence requires genuine feeling, understanding, or human-like inner experience that AI agents are unable to have. These perspectives clarify that some participants view social intelligence as requiring an inner experience, independent of an AI agent's observable behavior. 

\begin{figure}[b]
    \centering
    \includegraphics[width=\linewidth]{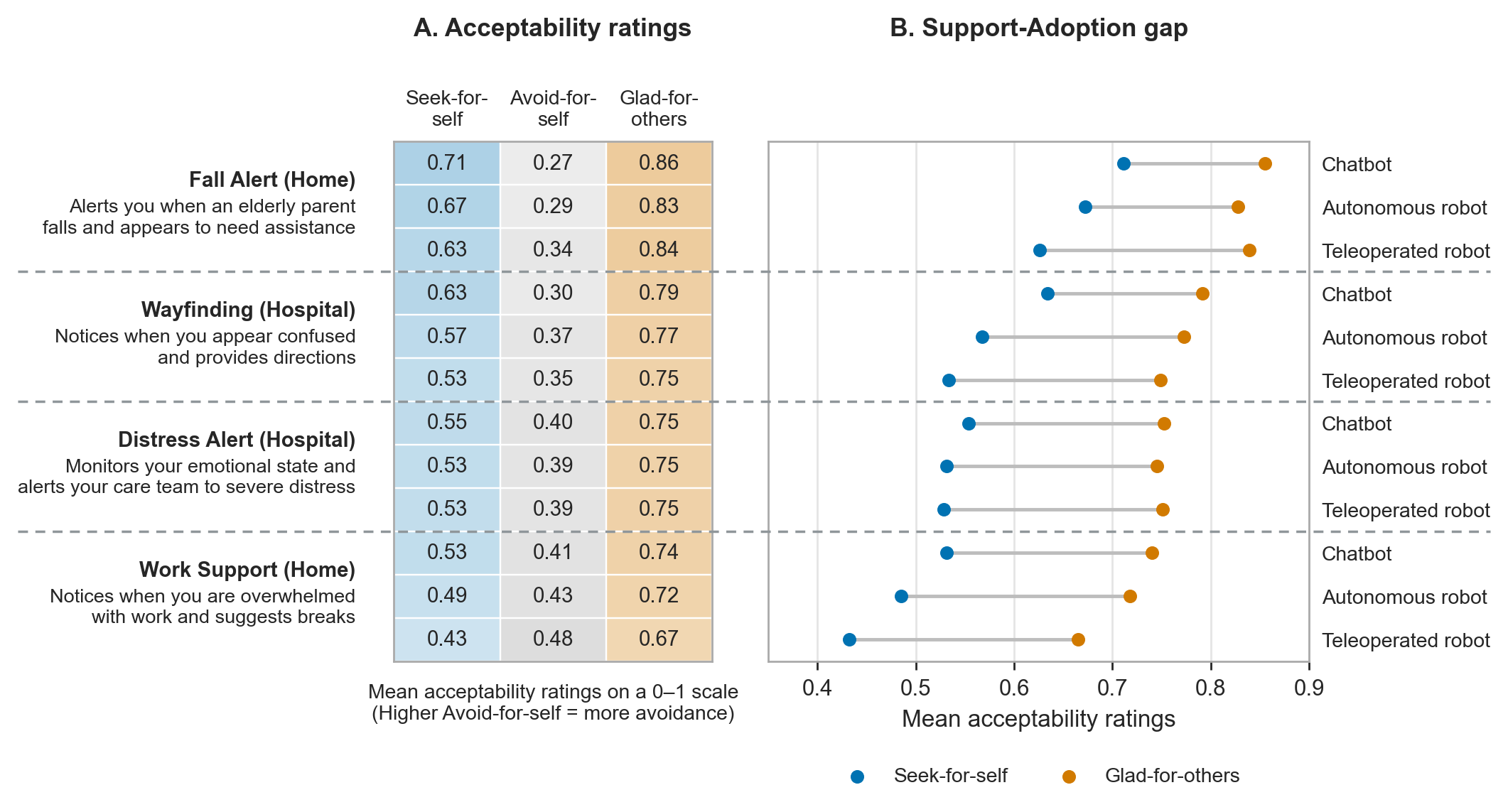}
    \caption{Acceptability ratings for the 12 scenario conditions from Table~\ref{tab:scenario-use-cases}. Panel A reports the Seek-for-self, Avoid-for-self, and Glad-for-others mean acceptability ratings for the 200 survey responses for each of the 12 scenario conditions. Panel B illustrates the Support-Adoption gap by comparing the Seek-for-self and Glad-for-others ratings for the same scenario context and agent type. Dashed lines separate the four deployment scenarios across both panels. All 12 paired Wilcoxon comparisons remained significant after Holm correction (adjusted $p$ < .001). 
    }
    \label{fig:rq3_acceptability_gap}
\end{figure}

\paragraph{\textbf{Implications of RQ2 Findings}:} \textit{We find that social intelligence in AI agents is perceived by laypeople as a collection of observable agent behaviors that span a range of interactional abilities, rather than through a single capability.} These findings support the need for researchers and developers to avoid conflating specific social skills with a broader competence in social intelligence \cite{weis2005social, soto2021taking}. \textit{Our findings imply that  Social-AI evaluation methods should distinguish between the  capabilities that an agent reliably demonstrates and the social intelligence  that users attribute to it, based upon its behavior.} Disentangling these two levels can help designers understand how interaction choices shape user perception of social intelligence in AI agents and avoid encouraging expectations beyond what the agent can deliver.

\subsection{Layperson Acceptance of Social-AI Agents}
\label{subsec:rq3}

To study how deployment context and agent type influence acceptability of Social-AI, we analyzed participants' ratings of the 12 scenario conditions summarized in Table \ref{tab:scenario-use-cases}. For each, participants rated whether they would seek out the service for themselves, avoid it, or be glad that it existed for others who wanted it. We analyzed these three ratings separately, normalizing each to a 0-1 scale. Higher Seek-for-self and Glad-for-others ratings indicate more favorable acceptability judgments about the Social-AI agent in given deployment scenarios, while higher Avoid-for-self ratings indicate greater inclination to avoid these agents.  

Acceptability ratings for the 12 scenario conditions are in Figure \ref{fig:rq3_acceptability_gap}, Panel A; means and 95\% confidence intervals are in Appendix Table \ref{tab:rq3_condition_means}.
Averaged across agent types, the \textit{Fall Alert (Home)} scenario received the highest Seek-for-self ratings and Glad-for-others ratings and the lowest Avoid-for-self ratings. \textit{Wayfinding (Hospital)}, \textit{Distress Alert (Hospital)}, and \textit{Work Support (Home)} followed, in that order. Participants were most willing to seek the chatbot \textit{Fall Alert} service (M = 0.71) and least willing to seek the teleoperated robot \textit{Work Support} service (M = 0.43).  Models accounting for repeated ratings from each participant indicate that Seek-for-self, Avoid-for-self, and Glad-for-others ratings each differed significantly across scenario contexts and across agent types (Appendix Table \ref{tab:rq3_omnibus}). 

\subsubsection{The Support-Adoption Gap}
\label{subsubsec:gap}
We discovered a significant Support-Adoption gap across all 12 scenarios; participants support the existence of Social-AI agents for others far more than for their own personal use. In Figure \ref{fig:rq3_acceptability_gap} Panel B, we visualize this phenomenon. Within each condition, we compared the two ratings from the same participants. Mean Support-Adoption gaps ranged from 0.14 to 0.23, with an overall mean difference of 0.20 points. All 12 Wilcoxon signed-rank tests remained significant after Holm correction (all adjusted $p$ < .001; Appendix Table \ref{tab:rq3_gap}), indicating that participants consistently supported the availability of Social-AI more than their willingness to personally adopt it. 

\subsubsection{AI Literacy and Acceptability Ratings of Social-AI Agents}
\label{subsubsec:literacy_acceptability}

\begin{figure}[t]
    \centering
    \includegraphics[width=\linewidth]{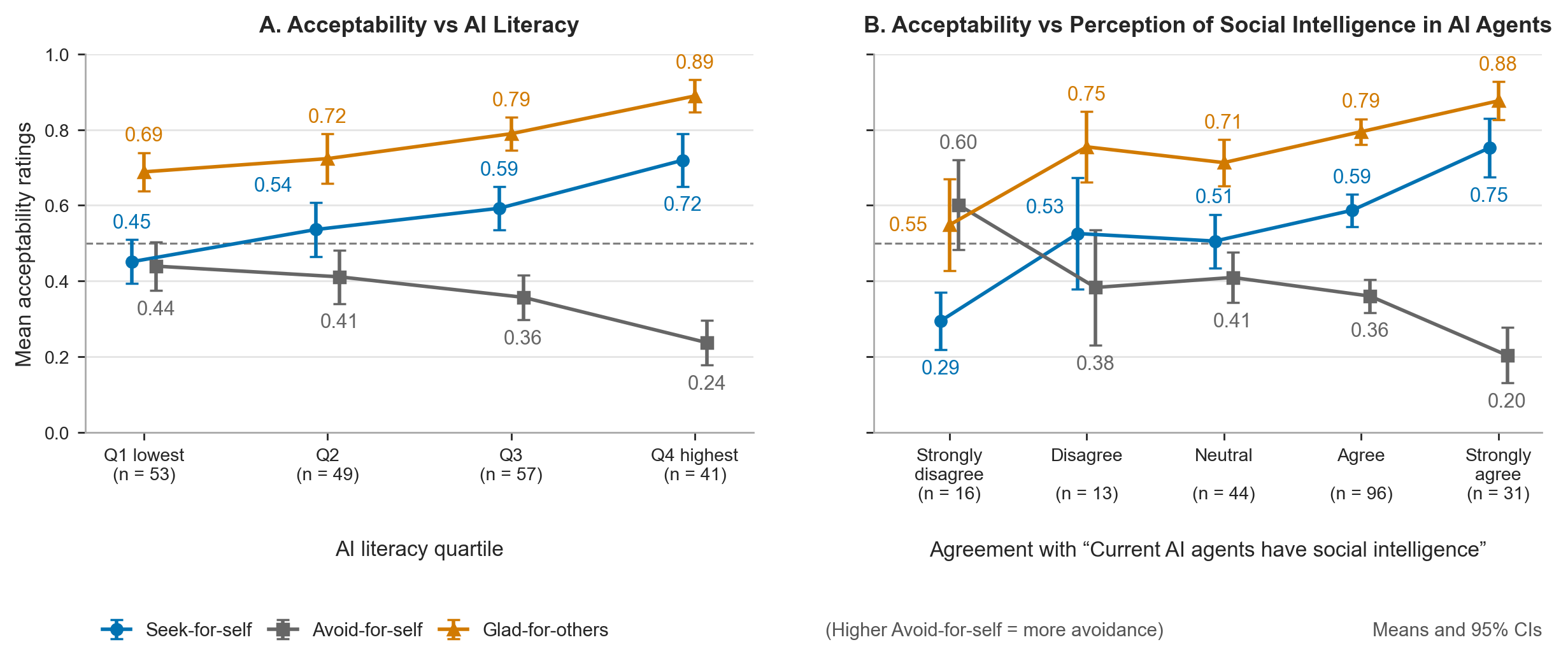}
    \caption{Panel A: Acceptability ratings for participants with different levels of AI literacy. Panel B: Acceptability ratings for participants with different agreement with the statement: ``Current AI agents have social intelligence.''}
    \label{fig:ai_literacy_acceptability}
\end{figure}

Higher self-reported AI literacy was associated with greater willingness to seek Social-AI services for oneself (Spearman's $\rho = .369$), lower avoidance ($\rho = -.293$), and higher ratings for being glad that the services existed for others ($\rho = .388$). All three associations remained significant after Benjamini--Hochberg correction (all adjusted $p < .001$; Appendix Table \ref{tab:rq3_secondary_correlations}). Figure \ref{fig:ai_literacy_acceptability}, Panel A, shows the ratings by AI-literacy quartile. From the lowest to the highest quartile, mean Seek-for-self ratings increased from 0.45 to 0.72, Avoid-for-self decreased from 0.44 to 0.24, and Glad-for-others increased from 0.69 to 0.89 (Appendix Table \ref{tab:rq3_predictor_group_means}). The associations between literacy and all three outcomes remained significant after accounting for reported chatbot and robot use frequency (all adjusted $p < .05$; Appendix Table \ref{tab:literacy_prior_use}).

\subsubsection{Social-AI Perceptions and Acceptability Ratings}
\label{subsubsec:perception_acceptability}

Participants who rated current AI agents as more socially intelligent on the single holistic item in Block B reported greater willingness to seek Social-AI services for themselves (Spearman's $\rho = .404$), lower avoidance ($\rho = -.348$), and higher ratings for being glad that the services existed for others ($\rho = .367$). All three associations remained significant after Benjamini--Hochberg correction (all adjusted $p < .001$; Appendix Table \ref{tab:rq3_secondary_correlations}). Figure \ref{fig:ai_literacy_acceptability}, Panel B, shows mean Seek-for-self, Avoid-for-self, and Glad-for-others ratings grouped by participants' agreement with the statement ``Current AI agents have social intelligence'' (Appendix Table \ref{tab:rq3_predictor_group_means}). These findings suggest that participants who perceive current AI agents as more socially intelligent are more willing to use Social-AI services themselves and more welcoming of their availability to others.

\paragraph{\textbf{Implications of RQ3 Findings:}}

\textit{We find that layperson acceptability ratings for Social-AI agents differ across deployment scenario contexts and agent types, and we find a consistent support-adoption gap across all  scenarios, with participants supporting the existence of Social-AI agents for others far more than for their own personal use}. \textit{These findings have public-facing implications for the design of future polling methods to assess public support for Social-AI.} Our research suggests that public surveys and impact assessments should separately assess broader support and personal adoption willingness, rather than treating the former as a proxy for the latter.  Surveying people to assess their level of support for the availability of a Social-AI agent can risk overstating their personal willingness to adopt the technology themselves.

\textit{We find that acceptability of Social-AI agents is positively associated with participants' self-reported AI literacy and their existing perceptions of AI agents having social intelligence.} \textit{These findings imply that participatory design and evaluation should proactively include people with diverse AI literacy and perceptions of AI}, rather than allowing those already receptive to the technology to signal acceptable deployment for everyone.

\subsection{Layperson Concerns about Social-AI Agents}
\label{subsec:concerns}

\subsubsection{Concerns and Reservations Extracted from Open-Ended Responses}
\label{subsubsec:concerns}

\begin{figure}[t]
    \centering
    \includegraphics[width=\linewidth]{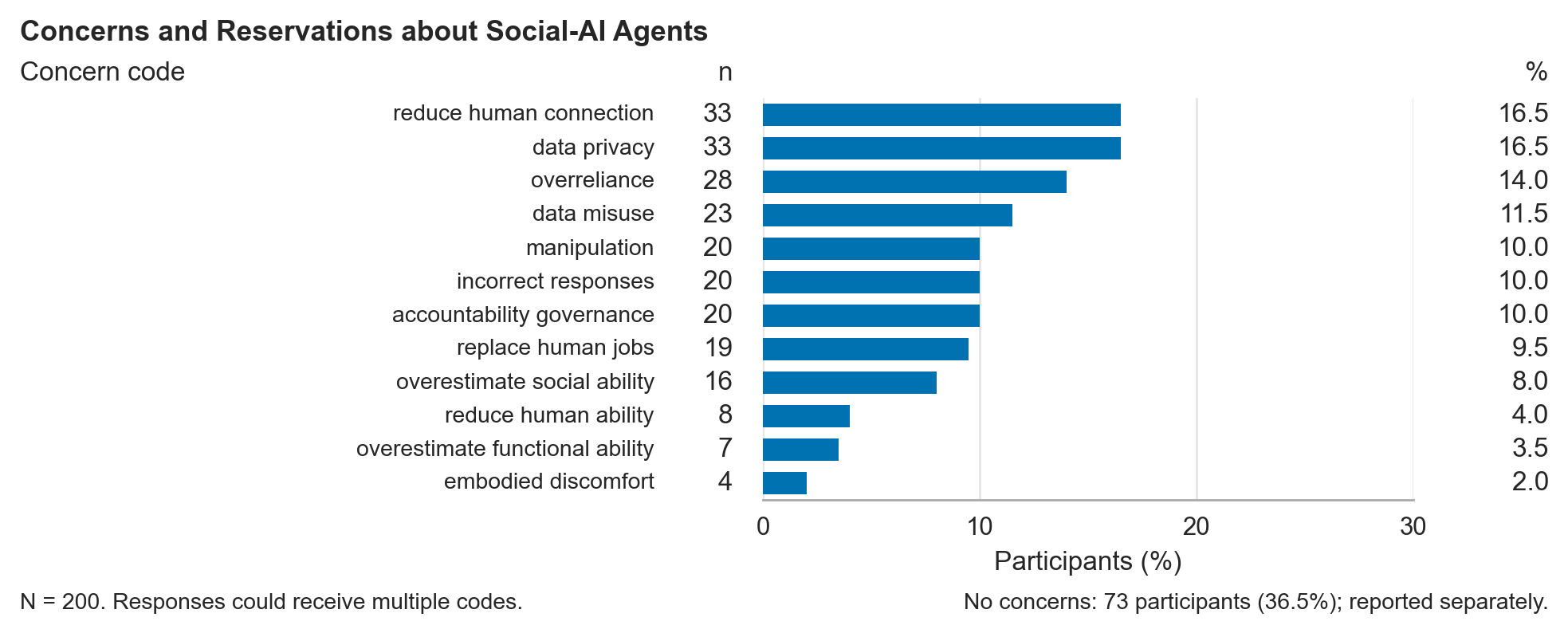}
    \caption{Distribution of RQ4 codes that participants identified as concerns or reservations about Social-AI agents. The graph shows code prevalence as percentages of the 200 participants. Responses could receive multiple codes.}
    \label{fig:rq4_concerns}
\end{figure}

To examine public concerns about Social-AI, we analyzed open-ended responses from Block D using the  coding process described in Section \ref{sec:methods}. We coded responses with an inductively developed multi-label codebook (Table \ref{tab:rq4_codebook}). Participants communicated a range of concerns and reservations, including social and psychological harms, governance and control, privacy and data risks, epistemic and functional risks, and labor or embodiment-related concerns. The most frequently occurring concerns raised by laypeople were related to reduced human connection (16.5\%), data privacy (16.5\%), user overreliance on the agent (14.0\%), data misuse (11.5\%),  accountability and governance (10.0\%), manipulation (10.0\%), and hallucinations and incorrect responses (10.0\%). The distribution of RQ4 codes, in order of prevalence, is in Figure \ref{fig:rq4_concerns}. 

While participants expressed a range of concerns and reservations about Social-AI, a substantial portion reported no concerns (73 participants; 36.5\%; Table \ref{tab:rq4_codebook}). Participants were explicitly asked to describe their concerns or write that they had none, so we interpret these responses as no concerns expressed in this survey. The no-concerns category also included positive views of Social-AI. For example, one participant wrote, ``I like that they exist and seem to be getting better all of the time. I look forward to a time when they can help even more'' (Appendix Table \ref{tab:rq4_representative_quotes_appendix}). These responses show that the concerns identified in our analysis were not shared by all participants, and that at least some participants explicitly welcomed Social-AI. We note that expressing no concerns when reasoning about this question does not establish acceptance of every deployment scenario.

\begin{table}
\centering
\small
\setlength{\tabcolsep}{2pt}
\renewcommand{\arraystretch}{1.05}
\begin{tabular}{p{0.17\linewidth}p{0.22\linewidth}p{0.45\linewidth}rr}
\hline
\textbf{Concern Cluster} & \textbf{Code} & \textbf{Definition} & \textbf{Count} & \textbf{\%} \\
\hline
No concerns & no concerns & Reports no concerns or reservations. & 73 & 36.5 \\
\hline
Social/psychological  & reduce human connection & Social-AI may reduce, replace, distort human connection. & 33 & 16.5 \\
Social/psychological  & overreliance & Users may become too dependent on Social-AI. & 28 & 14.0 \\
Social/psychological  & reduce human ability & Social-AI may weaken human abilities, skills, judgment. & 8 & 4.0 \\
\hline
Privacy/data & data privacy & Collection, exposure, or surveillance of personal data. & 33 & 16.5 \\
Privacy/data & data misuse & Collected data could be misused, hacked,  exploited. & 23 & 11.5 \\
\hline
Governance & accountability governance & Insufficient oversight, responsibility, or regulation. & 20 & 10.0 \\
Governance & manipulation & Social-AI may persuade, influence, or manipulate users. & 20 & 10.0 \\
\hline
Functional risks & incorrect responses & Social-AI may hallucinate information or mislead  users. & 20 & 10.0 \\
Functional risks & overestimate social ability & Users may overestimate the agent's social understanding. & 16 & 8.0 \\
Functional risks & overestimate functional ability & Users may overestimate the agent's task competence. & 7 & 3.5 \\
\hline
Labor/embodiment & replace human jobs & Concern that Social-AI may replace human workers. & 19 & 9.5 \\
Labor/embodiment & embodied discomfort & Discomfort related to embodied robots or physical malfunction. & 4 & 2.0 \\
\hline
\end{tabular}
\caption{RQ4 codes for concerns about Social-AI. Representative quotations are provided in the Appendix Table \ref{tab:rq4_representative_quotes_appendix}. Responses could receive multiple codes, so percentages do not sum to 100\%.}
\label{tab:rq4_codebook}
\end{table}

\subsubsection{Privacy and Data Preferences}
\label{subsubsec:privacy}


\begin{figure}
    \centering
    \includegraphics[width=0.5\linewidth]{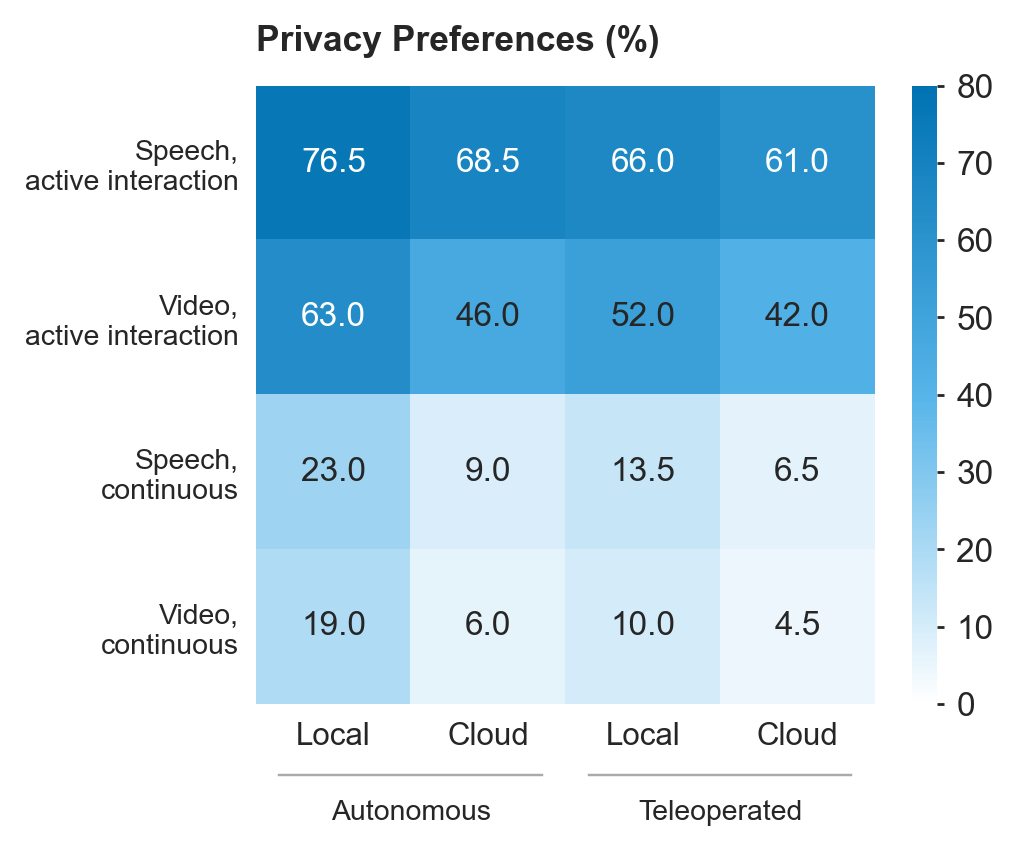}
    \caption{Participants' privacy preferences for Social-AI agents in homes. Percentage of participants selecting the combination of sensing modality/duration, data storage location, and robot autonomy.}
    \label{fig:rq3_privacy}
\end{figure}

We examined participants' preferences regarding the kinds of interaction data they would allow an embodied Social-AI agent at home to use and store. Each participant indicated their preferences under four conditions, varying (1) whether the robot was autonomous or teleoperated and (2) whether interaction data was stored locally or in the cloud. In each case, participants could select  information they would be willing to have the robot use and store: continuous speech, speech only during active interaction, continuous video, video only during active interaction, or none of the above. Active interaction sensing occurs while the user is directly interacting with the  robot. Participants' recorded selections are visualized in Figure \ref{fig:rq3_privacy} and reported in Appendix Table \ref{tab:privacy_preferences_appendix}.

Participants more frequently selected sensing during active interaction than continuous sensing. Across the four robot and storage conditions, active-interaction speech and video options were selected by 42.0\%--76.5\% of participants, compared with 4.5\%--23.0\% for continuous sensing. All eight comparisons of active-interaction and continuous sensing within the same participants were significant after Holm correction (exact McNemar tests, all adjusted $p < .001$; Appendix Table \ref{tab:privacy_paired_tests}). Speech options were selected more frequently than video options, local storage more frequently than cloud storage, and autonomous robots more frequently than teleoperated robots in every matched comparison. The most frequently selected option was 
to permit data use by autonomous agents using speech sensing (76.5\%)  during direct interaction with data stored locally, and the least frequently selected option was continuous video sensing by a teleoperated robot with cloud storage (4.5\%). The next lowest were continuous video sensing by an autonomous robot with cloud storage (6.0\%), continuous speech sensing by a teleoperated robot with cloud storage (6.5\%), and continuous speech sensing by an autonomous robot with cloud storage (9.0\%).

\paragraph{\textbf{Implications of RQ4 Findings}:} 
\textit{We find that participants express a range of concerns about Social-AI agents; primary concerns include reduced human connection and emotional and cognitive overreliance on agents, alongside harms arising from privacy issues, data misuse, and manipulation.} Prior studies and frameworks for AI risks have identified overreliance and emotional dependence as potential harms of AI systems \cite{ai2024artificial, slattery2024ai, brauner2026charting}. Our findings complement 
prior works by contributing empirical evidence that these  concerns are especially salient in layperson perceptions of Social-AI. For researchers and developers, these concerns motivate designing Social-AI agents that support and augment human social connection and cognitive abilities \cite{jiang2026scaffolded, xu2025productive}. Our findings highlight the need for future longitudinal HCI studies to examine how Social-AI system design can address user concerns about these technologies. 

\textit{We find that participants' willingness to allow Social-AI agents to use and store data in their homes varied depending on what modalities were sensed by agents, whether or not sensing was continuous, where data were stored, and whether the agent was autonomous or teleoperated. Participants were more willing to permit sensing of speech data (rather than video), 
during direct interaction with AI agents (rather than continuous sensing), with data being stored locally (rather than on the cloud), and by autonomous agents (rather than teleoperated agents with humans in-the-loop).} Our 
findings uncover the need for researchers and developers to create Social-AI agents that give users authority over which modalities are sensed, when sensing is active, and where interaction data are stored \cite{stapels2023never}. 
The preference expressed by participants for local storage over cloud storage motivates the exploration of \textit{on-device social perception and reasoning algorithms} that can reduce the need to transmit sensitive interaction data to cloud services \cite{mathur-etal-2024-advancing}. Participants' lower willingness to permit data use by teleoperated Social-AI agents has implications for ongoing efforts by AI researchers and developers to scale robot learning through the collection of teleoperated demonstrations \cite{butler2015privacy, parts2025systematic}. Our findings motivate future  research that examines how to design, deploy, and evaluate Social-AI agents that maintain technical performance, while respecting user preferences and boundaries.

\section{Conclusions}
\label{sec:governance}

As socially intelligent AI agents increasingly move from labs into everyday settings such as homes and hospitals, understanding public perception and acceptance of these agents is essential to inform responsible and participatory design, development, and deployment of these technologies to support human well-being. Our paper contributes new empirical knowledge about layperson perceptions of social intelligence in current AI agents, AI agent behaviors that laypeople associate with social intelligence, layperson acceptance of Social-AI agents across deployment scenarios, and layperson concerns and reservations about Social-AI. We find that social intelligence in AI is not merely a future technical goal imagined by researchers, but already a part of how laypeople view the AI agents they encounter. We uncovered a \textit{support-adoption} gap, with participants more willing to support the availability of Social-AI agents for others than to personally adopt these systems for themselves. Our findings reveal that participants recognize potential benefits of Social-AI, but hold reservations about multiple dimensions of risk related to privacy, data misuse, and social and psychological harms. Our findings regarding laypersons' lower willingness to permit continuous sensing of interaction data by a Social-AI agent, storage of their interaction data in the cloud, and the use of teleoperated agents in their spaces (instead of autonomous agents) inform the design, development, and deployment of future Social-AI agents that are likely to be responsive to user preferences and accepted by the public. 

\section*{Acknowledgments}
Leena Mathur acknowledges support from the SoftBank Group-Arm Fellowship and the NSF Graduate Research Fellowship Program under Grant No. DGE2140739.  Maarten Sap acknowledges support from Google, in part, for Prolific data collection funding. Opinions, findings, conclusions, or recommendations expressed in this material are those of the authors and
do not necessarily reflect the views of the sponsors, and no official endorsement should be inferred.

\bibliographystyle{ACM-Reference-Format}
\bibliography{sample-base}

@inproceedings{mathur-etal-2024-advancing,
    title = "Advancing Social Intelligence in {AI} Agents: Technical Challenges and Open Questions",
    author = "Mathur, Leena  and
      Liang, Paul Pu  and
      Morency, Louis-Philippe",
    editor = "Al-Onaizan, Yaser  and
      Bansal, Mohit  and
      Chen, Yun-Nung",
    booktitle = "Proceedings of the 2024 Conference on Empirical Methods in Natural Language Processing",
    month = nov,
    year = "2024",
    address = "Miami, Florida, USA",
    publisher = "Association for Computational Linguistics",
    url = "https://aclanthology.org/2024.emnlp-main.1143/",
    doi = "10.18653/v1/2024.emnlp-main.1143",
    pages = "20541--20560"
}

@inproceedings{bondi2021envisioning,
  title={Envisioning communities: a participatory approach towards AI for social good},
  author={Bondi, Elizabeth and Xu, Lily and Acosta-Navas, Diana and Killian, Jackson A},
  booktitle={Proceedings of the 2021 AAAI/ACM Conference on AI, Ethics, and Society},
  pages={425--436},
  year={2021}
}

@inproceedings{li2024social,
  title={Social intelligence data infrastructure: Structuring the present and navigating the future},
  author={Li, Minzhi and Shi, Weiyan and Ziems, Caleb and Yang, Diyi},
  booktitle={Findings of the Association for Computational Linguistics: ACL 2024},
  pages={2789--2805},
  year={2024}
}

@article{nass2000machines,
  title={Machines and mindlessness: Social responses to computers},
  author={Nass, Clifford and Moon, Youngme},
  journal={Journal of social issues},
  volume={56},
  number={1},
  pages={81--103},
  year={2000},
  publisher={Wiley Online Library}
}

@article{thorndike1937evaluation,
  title={An evaluation of the attempts to measure social intelligence.},
  author={Thorndike, Robert L and Stein, Saul},
  journal={Psychological bulletin},
  volume={34},
  number={5},
  pages={275},
  year={1937},
  publisher={Psychological Review Company}
}

@article{weis2005social,
  title={Social intelligence--A review and critical discussion of measurement concepts},
  author={Weis, Susanne and S{\"u}{\ss}, Heinz-Martin},
  journal={Emotional intelligence: An international handbook},
  pages={203--230},
  year={2005},
  publisher={Hogrefe \& Huber: Washington, USA}
}

@article{dennler2023design,
  title={Design metaphors for understanding user expectations of socially interactive robot embodiments},
  author={Dennler, Nathaniel and Ruan, Changxiao and Hadiwijoyo, Jessica and Chen, Brenna and Nikolaidis, Stefanos and Matari{\'c}, Maja},
  journal={ACM Transactions on Human-Robot Interaction},
  volume={12},
  number={2},
  pages={1--41},
  year={2023},
  publisher={ACM New York, NY}
}

@article{nakanishi2020continuous,
  title={Continuous hospitality with social robots at a hotel},
  author={Nakanishi, Junya and Kuramoto, Itaru and Baba, Jun and Ogawa, Kohei and Yoshikawa, Yuichiro and Ishiguro, Hiroshi},
  journal={SN Applied Sciences},
  volume={2},
  number={3},
  pages={452},
  year={2020},
  publisher={Springer}
}

@inproceedings{ingber2025distinguishing,
  title={Distinguishing Emotion AI: Factors Shaping Perceptions Including Input Data, Emotion Data Recipients, and Identity},
  author={Ingber, Alexis Shore and Haimson, Oliver L and Andalibi, Nazanin},
  booktitle={Proceedings of the 2025 ACM Conference on Fairness, Accountability, and Transparency},
  pages={498--510},
  year={2025}
}

@inproceedings{andalibi2025public,
  title={Public Perceptions About Emotion AI Use Across Contexts in the United States},
  author={Andalibi, Nazanin and Ingber, Alexis Shore},
  booktitle={Proceedings of the 2025 CHI Conference on Human Factors in Computing Systems},
  pages={1--16},
  year={2025}
}

@misc{heerink2010assessing,
  title={Assessing acceptance of assistive social agent technology by older adults: the almere model},
  author={Heerink, Marcel and Kr{\"o}se, Ben and Evers, Vanessa and Wielinga, Bob},
  year={2010},
  journal = {International Journal of Social Robotics},
  publisher={Springer}
}

@inproceedings{takayama2008beyond,
  title={Beyond dirty, dangerous and dull: what everyday people think robots should do},
  author={Takayama, Leila and Ju, Wendy and Nass, Clifford},
  booktitle={Proceedings of the 3rd ACM/IEEE international conference on Human robot interaction},
  pages={25--32},
  year={2008}
}

@article{weizenbaum1983eliza,
  title={ELIZA—a computer program for the study of natural language communication between man and machine},
  author={Weizenbaum, Joseph},
  journal={Communications of the ACM},
  volume={26},
  number={1},
  pages={23--28},
  year={1983},
  publisher={ACM New York, NY, USA}
}

@article{natale2019if,
  title={If software is narrative: Joseph Weizenbaum, artificial intelligence and the biographies of ELIZA},
  author={Natale, Simone},
  journal={New Media \& Society},
  volume={21},
  number={3},
  pages={712--728},
  year={2019},
  publisher={SAGE Publications Sage UK: London, England}
}

@inproceedings{brown2024trash,
  title={Trash in motion: Emergent interactions with a robotic trashcan},
  author={Brown, Barry and Bu, Fanjun and Mandel, Ilan and Ju, Wendy},
  booktitle={Proceedings of the 2024 CHI Conference on Human Factors in Computing Systems},
  pages={1--17},
  year={2024}
}

@article{henschel2021makes,
  title={What makes a robot social? A review of social robots from science fiction to a home or hospital near you},
  author={Henschel, Anna and Laban, Guy and Cross, Emily S},
  journal={Current Robotics Reports},
  volume={2},
  number={1},
  pages={9--19},
  year={2021},
  publisher={Springer}
}

@article{hurst2020social,
  title={Social and emotional skills training with embodied moxie},
  author={Hurst, Nikki and Clabaugh, Caitlyn and Baynes, Rachel and Cohn, Jeff and Mitroff, Donna and Scherer, Stefan},
  journal={arXiv preprint arXiv:2004.12962},
  year={2020}
}

@online{handa2026enablingindependentresearch,
author = {Kunal Handa and Miranda Zhang and Gabriel Nicholas and Miles McCain and Ryan Heller and Saffron Huang and Thomas Millar and Suzanne Wang and Shan Carter and Mo Julapalli and Matt Kearney and Sarah Pollack and Judy Shen and Matthew Jagielski and Shaoyi Zhang and Heather Whitney and Ankur Rathi and Aisling Keenan and David Saunders and Jake Eaton and Sylvie Carr and Jack Clark and Michael Stern and Deep Ganguli},
title = {Enabling independent research on how people use Claude},
date = {2026-08-26},
year = {2026},
url = {www.anthropic.com/research/enabling-independent-research},
}

@inproceedings{ranganeni2024robots,
  title={Robots for humanity: in-home deployment of Stretch RE2},
  author={Ranganeni, Vinitha and Nguyen, Vy and Evans, Henry and Evans, Jane and Mehu, Julian and Olatunji, Samuel and Rogers, Wendy and Edsinger, Aaron and Kemp, Charles and Cakmak, Maya},
  booktitle={Companion of the 2024 ACM/IEEE international conference on human-robot interaction},
  pages={1299--1301},
  year={2024}
}

@book{10.7551/mitpress/4398.001.0001,
    author = {Carroll, John M.},
    title = {Making Use: Scenario-Based Design of Human-Computer Interactions},
    publisher = {The MIT Press},
    year = {2000},
    month = {08},
    isbn = {9780262269926},
    doi = {10.7551/mitpress/4398.001.0001},
    url = {https://doi.org/10.7551/mitpress/4398.001.0001},
}

@techreport{NBERw34255,
 title = "How People Use ChatGPT",
 author = "Chatterji, Aaron and Cunningham, Thomas and Deming, David J and Hitzig, Zoe and Ong, Christopher and Shan, Carl Yan and Wadman, Kevin",
 institution = "National Bureau of Economic Research",
 type = "Working Paper",
 series = "Working Paper Series",
 number = "34255",
 year = "2025",
 month = "September",
 doi = {10.3386/w34255},
 URL = "http://www.nber.org/papers/w34255",
}

@article{blut2021understanding,
  title={Understanding anthropomorphism in service provision: a meta-analysis of physical robots, chatbots, and other AI},
  author={Blut, Markus and Wang, Cheng and W{\"u}nderlich, Nancy V and Brock, Christian},
  journal={Journal of the academy of marketing science},
  volume={49},
  number={4},
  pages={632--658},
  year={2021},
  publisher={Springer}
}

@inproceedings{10.1145/642611.642635,
author = {Palen, Leysia and Dourish, Paul},
title = {Unpacking "privacy" for a networked world},
year = {2003},
isbn = {1581136307},
publisher = {Association for Computing Machinery},
address = {New York, NY, USA},
url = {https://doi.org/10.1145/642611.642635},
doi = {10.1145/642611.642635},
booktitle = {Proceedings of the SIGCHI Conference on Human Factors in Computing Systems},
pages = {129–136},
numpages = {8},
location = {Ft. Lauderdale, Florida, USA},
series = {CHI '03}
}

@inproceedings{thomaz2023robots,
  title={Robots in real life: Putting hri to work},
  author={Thomaz, Andrea},
  booktitle={Proceedings of the 2023 ACM/IEEE International Conference on Human-Robot Interaction},
  pages={3--3},
  year={2023}
}

@inproceedings{xu2025productive,
  title={Productive vs. reflective: how different ways of integrating AI into design workflows affect cognition and motivation},
  author={Xu, Xiaotong and Konnova, Arina and Gao, Bianca and Peng, Cindy and Vo, Dave and Dow, Steven P},
  booktitle={Proceedings of the 2025 CHI Conference on Human Factors in Computing Systems},
  pages={1--15},
  year={2025}
}

@article{stapels2023never,
  title={Never trust anything that can think for itself, if you can’t control its privacy settings: The influence of a robot’s privacy settings on users’ attitudes and willingness to self-disclose},
  author={Stapels, Julia G and Penner, Angelika and Diekmann, Niels and Eyssel, Friederike},
  journal={International Journal of Social Robotics},
  volume={15},
  number={9},
  pages={1487--1505},
  year={2023},
  publisher={Springer}
}

@article{barchard2020measuring,
  title={Measuring the perceived social intelligence of robots},
  author={Barchard, Kimberly A and Lapping-Carr, Leiszle and Westfall, R Shane and Fink-Armold, Andrea and Banisetty, Santosh Balajee and Feil-Seifer, David},
  journal={ACM Transactions on Human-Robot Interaction (THRI)},
  volume={9},
  number={4},
  pages={1--29},
  year={2020},
  publisher={ACM New York, NY, USA}
}

@inproceedings{10.1145/3772318.3790936,
author = {Wang, Qiaosi and Kim, Jini and Sharma, Avanita and Lee, Alicia (Hyun Jin) and Forlizzi, Jodi and Shen, Hong},
title = {Situated, Dynamic, and Subjective: Envisioning the Design of Theory-of-Mind-Enabled Everyday AI with Industry Practitioners},
year = {2026},
isbn = {9798400722783},
publisher = {Association for Computing Machinery},
address = {New York, NY, USA},
url = {https://doi.org/10.1145/3772318.3790936},
doi = {10.1145/3772318.3790936},
booktitle = {Proceedings of the 2026 CHI Conference on Human Factors in Computing Systems},
articleno = {1379},
numpages = {21},
location = {
},
series = {CHI '26}
}

@article{brinkschulte2022perspectives,
  title={Perspectives on Socially Intelligent Conversational Agents},
  author={Brinkschulte, Luisa and Schl{\"o}gl, Stephan and Monz, Alexander and Sch{\"o}ttle, Pascal and Janetschek, Matthias},
  journal={Multimodal Technologies and Interaction},
  volume={6},
  number={8},
  pages={62},
  year={2022},
  publisher={MDPI}
}

@inproceedings{butler2015privacy,
  title={The privacy-utility tradeoff for remotely teleoperated robots},
  author={Butler, Daniel J and Huang, Justin and Roesner, Franziska and Cakmak, Maya},
  booktitle={Proceedings of the tenth annual ACM/IEEE international conference on human-robot interaction},
  pages={27--34},
  year={2015}
}

@inproceedings{jiang2026scaffolded,
  title={Scaffolded Vulnerability: Chatbot-Mediated Reciprocal Self-Disclosure and Need-Supportive Interaction in Couples},
  author={Jiang, Zhuoqun and Yeo, ShunYi and Herremans, Dorien and Tangi Perrault, Simon},
  booktitle={Proceedings of the 2026 CHI Conference on Human Factors in Computing Systems},
  pages={1--39},
  year={2026}
}

@article{soto2021taking,
  title={Taking skills seriously: Toward an integrative model and agenda for social, emotional, and behavioral skills},
  author={Soto, Christopher J and Napolitano, Christopher M and Roberts, Brent W},
  journal={Current Directions in Psychological Science},
  volume={30},
  number={1},
  pages={26--33},
  year={2021},
  publisher={Sage Publications Sage CA: Los Angeles, CA}
}

@article{wang2025rethinking,
  title={Rethinking theory of mind benchmarks for llms: Towards a user-centered perspective},
  author={Wang, Qiaosi and Zhou, Xuhui and Sap, Maarten and Forlizzi, Jodi and Shen, Hong},
  journal={arXiv preprint arXiv:2504.10839},
  year={2025}
}

@article{benjamini1995controlling,
  title={Controlling the false discovery rate: a practical and powerful approach to multiple testing},
  author={Benjamini, Yoav and Hochberg, Yosef},
  journal={Journal of the Royal statistical society: series B (Methodological)},
  volume={57},
  number={1},
  pages={289--300},
  year={1995},
  publisher={Wiley Online Library}
}

@inproceedings{10.1145/3290605.3300233,
author = {Amershi, Saleema and Weld, Dan and Vorvoreanu, Mihaela and Fourney, Adam and Nushi, Besmira and Collisson, Penny and Suh, Jina and Iqbal, Shamsi and Bennett, Paul N. and Inkpen, Kori and Teevan, Jaime and Kikin-Gil, Ruth and Horvitz, Eric},
title = {Guidelines for Human-AI Interaction},
year = {2019},
isbn = {9781450359702},
publisher = {Association for Computing Machinery},
address = {New York, NY, USA},
url = {https://doi.org/10.1145/3290605.3300233},
doi = {10.1145/3290605.3300233},
booktitle = {Proceedings of the 2019 CHI Conference on Human Factors in Computing Systems},
pages = {1–13},
numpages = {13},
location = {Glasgow, Scotland Uk},
series = {CHI '19}
}

@article{deng2019embodiment,
  title={Embodiment in socially interactive robots},
  author={Deng, Eric and Mutlu, Bilge and Matari{\'c}, Maja J},
  journal={Foundations and Trends{\textregistered} in Robotics},
  volume={7},
  number={4},
  pages={251--356},
  year={2019},
  publisher={Emerald Publishing Limited}
}

@article{wilcoxon1945individual,
  title={Individual comparisons by ranking methods},
  author={Wilcoxon, Frank},
  journal={Biometrics bulletin},
  volume={1},
  number={6},
  pages={80--83},
  year={1945},
  publisher={JSTOR}
}

@article{spearman1961proof,
  title={The proof and measurement of association between two things.},
  author={Spearman, Charles},
  year={1961},
  publisher={Appleton-Century-Crofts}
}

@inproceedings{mushkani2025co,
  title={Co-Producing AI: Toward an Augmented, Participatory Lifecycle},
  author={Mushkani, Rashid and Berard, Hugo and Ammar, Toumadher and Chatonnier, Cassandre and Koseki, Shin},
  booktitle={Proceedings of the AAAI/ACM Conference on AI, Ethics, and Society},
  volume={8},
  number={2},
  pages={1785--1799},
  year={2025}
}

@inproceedings{ullstein2025participatory,
  title={Participatory AI and the EU AI Act},
  author={Ullstein, Chiara and Jarvers, Simon and Hohendanner, Michel and Papakyriakopoulos, Orestis and Grossklags, Jens},
  booktitle={Proceedings of the AAAI/ACM Conference on AI, Ethics, and Society},
  volume={8},
  number={3},
  pages={2550--2562},
  year={2025}
}

@inproceedings{mun2025not,
  title={Why (Not) Use AI? Analyzing People’s Reasoning and Conditions for AI Acceptability},
  author={Mun, Jimin and Yeong, Wei Bin Au and Deng, Wesley Hanwen and Borg, Jana Schaich and Sap, Maarten},
  booktitle={Proceedings of the AAAI/ACM Conference on AI, Ethics, and Society},
  volume={8},
  number={2},
  pages={1771--1784},
  year={2025}
}

@inproceedings{sap2019social,
  title={Social IQa: Commonsense reasoning about social interactions},
  author={Sap, Maarten and Rashkin, Hannah and Chen, Derek and Le Bras, Ronan and Choi, Yejin},
  booktitle={Proceedings of the 2019 conference on empirical methods in natural language processing and the 9th international joint conference on natural language processing (EMNLP-IJCNLP)},
  pages={4463--4473},
  year={2019}
}

@inproceedings{
zhou2024sotopia,
title={{SOTOPIA}: Interactive Evaluation for Social Intelligence in Language Agents},
author={Xuhui Zhou and Hao Zhu and Leena Mathur and Ruohong Zhang and Haofei Yu and Zhengyang Qi and Louis-Philippe Morency and Yonatan Bisk and Daniel Fried and Graham Neubig and Maarten Sap},
booktitle={The Twelfth International Conference on Learning Representations},
year={2024},
url={https://openreview.net/forum?id=mM7VurbA4r}
}

@article{corbin1990grounded,
  title={Grounded theory research: Procedures, canons, and evaluative criteria},
  author={Corbin, Juliet M and Strauss, Anselm},
  journal={Qualitative sociology},
  volume={13},
  number={1},
  pages={3--21},
  year={1990},
  publisher={Springer}
}

@inproceedings{mathur-etal-2025-social,
    title = "Social Genome: Grounded Social Reasoning Abilities of Multimodal Models",
    author = "Mathur, Leena  and
      Qian, Marian  and
      Liang, Paul Pu  and
      Morency, Louis-Philippe",
    editor = "Christodoulopoulos, Christos  and
      Chakraborty, Tanmoy  and
      Rose, Carolyn  and
      Peng, Violet",
    booktitle = "Proceedings of the 2025 Conference on Empirical Methods in Natural Language Processing",
    month = nov,
    year = "2025",
    address = "Suzhou, China",
    publisher = "Association for Computational Linguistics",
    url = "https://aclanthology.org/2025.emnlp-main.1264/",
    doi = "10.18653/v1/2025.emnlp-main.1264",
    pages = "24868--24891",
    ISBN = "979-8-89176-332-6"
}

@book{turner1988theory,
  title={A theory of social interaction},
  author={Turner, Jonathan H},
  year={1988},
  publisher={Stanford University Press}
}

@article{conzelmann2013new,
  title={New findings about social intelligence},
  author={Conzelmann, Kristin and Weis, Susanne and S{\"u}{\ss}, Heinz-Martin},
  journal={Journal of Individual Differences},
  year={2013},
  publisher={Hogrefe Publishing}
}

@inbook{Kihlstrom_Cantor_2000, place={Cambridge}, title={Social Intelligence}, booktitle={Handbook of Intelligence}, publisher={Cambridge University Press}, author={Kihlstrom, John F. and Cantor, Nancy}, editor={Sternberg, Robert J.Editor}, year={2000}, pages={359–379}}

@article{kaminski2016averting,
  title={Averting robot eyes},
  author={Kaminski, Margot E and Rueben, Matthew and Smart, William D and Grimm, Cindy M},
  journal={Md. L. Rev.},
  volume={76},
  pages={983},
  year={2016},
  publisher={HeinOnline}
}

@article{slattery2024ai,
  title={The ai risk repository: A comprehensive meta-review, database, and taxonomy of risks from artificial intelligence},
  author={Slattery, Peter and Saeri, Alexander K and Grundy, Emily AC and Graham, Jess and Noetel, Michael and Uuk, Risto and Dao, James and Pour, Soroush and Casper, Stephen and Thompson, Neil},
  journal={arXiv preprint arXiv:2408.12622},
  year={2024}
}

@article{ai2024artificial,
  title={Artificial intelligence risk management framework: Generative artificial intelligence profile},
  author={AI, NIST},
  journal={NIST Trustworthy and Responsible AI Gaithersburg, MD, USA},
  year={2024}
}

@article{wang2023measuring,
  title={Measuring user competence in using artificial intelligence: validity and reliability of artificial intelligence literacy scale},
  author={Wang, Bingcheng and Rau, Pei-Luen Patrick and Yuan, Tianyi},
  journal={Behaviour \& information technology},
  volume={42},
  number={9},
  pages={1324--1337},
  year={2023},
  publisher={Taylor \& Francis}
}

@inproceedings{selbst2019fairness,
  title={Fairness and abstraction in sociotechnical systems},
  author={Selbst, Andrew D and Boyd, Danah and Friedler, Sorelle A and Venkatasubramanian, Suresh and Vertesi, Janet},
  booktitle={Proceedings of the conference on fairness, accountability, and transparency},
  pages={59--68},
  year={2019}
}

@inproceedings{chen2026presenting,
  title={Presenting Large Language Models as Companions Affects What Mental Capacities People Attribute to Them},
  author={Chen, Allison and Kim, Sunnie SY and Franyutti-Cintron, Angel Nathaniel and Dharmasiri, Amaya and Mukherjee, Kushin and Russakovsky, Olga and Fan, Judith E},
  booktitle={Proceedings of the 2026 CHI Conference on Human Factors in Computing Systems},
  pages={1--30},
  year={2026}
}

@inproceedings{abercrombie2023mirages,
  title={Mirages. on anthropomorphism in dialogue systems},
  author={Abercrombie, Gavin and Curry, Amanda Cercas and Dinkar, Tanvi and Rieser, Verena and Talat, Zeerak},
  booktitle={Proceedings of the 2023 Conference on Empirical Methods in Natural Language Processing},
  pages={4776--4790},
  year={2023}
}

@article{parts2025systematic,
  title={A Systematic Scoping Review of Privacy Challenges and Privacy Enhancing Technologies in Teleoperated Robotics},
  author={Parts, Joosep and Leoste, Janika and Tammem{\"a}e, Kalle and Rakic, Slavko},
  journal={IEEE Access},
  volume={13},
  pages={216724--216747},
  year={2025},
  publisher={IEEE}
}

@inproceedings{paepcke2010judging,
  title={Judging a bot by its cover: An experiment on expectation setting for personal robots},
  author={Paepcke, Steffi and Takayama, Leila},
  booktitle={2010 5th ACM/IEEE international Conference on human-robot interaction (HRI)},
  pages={45--52},
  year={2010},
  organization={IEEE}
}

@article{gambino2020building,
  title={Building a stronger CASA: Extending the computers are social actors paradigm},
  author={Gambino, Andrew and Fox, Jesse and Ratan, Rabindra A},
  journal={Human-Machine Communication},
  volume={1},
  pages={71--85},
  year={2020},
  publisher={Communication and Social Robotics Labs Kalamazoo, Michigan}
}

@inproceedings{gordon2016affective,
  title={Affective personalization of a social robot tutor for children’s second language skills},
  author={Gordon, Goren and Spaulding, Samuel and Westlund, Jacqueline Kory and Lee, Jin Joo and Plummer, Luke and Martinez, Marayna and Das, Madhurima and Breazeal, Cynthia},
  booktitle={Proceedings of the AAAI conference on artificial intelligence},
  volume={30},
  number={1},
  year={2016}
}

@article{spitale2025past,
  title={Past, present, and future: A survey of the evolution of affective robotics for well-being},
  author={Spitale, Micol and Axelsson, Minja and Jeong, Sooyeon and Tutt{\"o}s{\'\i}, Paige and Stamatis, Caitlin A and Laban, Guy and Lim, Angelica and Gunes, Hatice},
  journal={IEEE Transactions on Affective Computing},
  year={2025},
  publisher={IEEE}
}

@misc{bls_soc_2018,
  author       = {{U.S. Bureau of Labor Statistics}},
  title        = {The 2018 Standard Occupational Classification System},
  year         = {2018},
  url          = {https://www.bls.gov/soc/},
}

@inproceedings{delgado2023participatory,
  title={The participatory turn in ai design: Theoretical foundations and the current state of practice},
  author={Delgado, Fernando and Yang, Stephen and Madaio, Michael and Yang, Qian},
  booktitle={Proceedings of the 3rd ACM Conference on Equity and Access in Algorithms, Mechanisms, and Optimization},
  pages={1--23},
  year={2023}
}

@article{gweon2023socially,
  title={Socially intelligent machines that learn from humans and help humans learn},
  author={Gweon, Hyowon and Fan, Judith and Kim, Been},
  journal={Philosophical Transactions of the Royal Society A: Mathematical, Physical and Engineering Sciences},
  volume={381},
  number={2251},
  year={2023},
  publisher={The Royal Society}
}

@article{fei2022searching,
  title={Searching for computer vision north stars},
  author={Fei-Fei, Li and Krishna, Ranjay},
  journal={Daedalus},
  volume={151},
  number={2},
  pages={85--99},
  year={2022},
  publisher={MIT Press One Rogers Street, Cambridge, MA 02142-1209, USA journals-info~…}
}

@incollection{krosnick2017questionnaire,
  title={Questionnaire design},
  author={Krosnick, Jon A},
  booktitle={The Palgrave handbook of survey research},
  pages={439--455},
  year={2017},
  publisher={Springer}
}

@book{marsden2010handbook,
  title={Handbook of survey research},
  author={Marsden, Peter V and Wright, James D},
  year={2010},
  publisher={Emerald Group Publishing}
}

@article{dunn2014alpha,
  title={From alpha to omega: A practical solution to the pervasive problem of internal consistency estimation},
  author={Dunn, Thomas J and Baguley, Thom and Brunsden, Vivienne},
  journal={British journal of psychology},
  volume={105},
  number={3},
  pages={399--412},
  year={2014},
  publisher={Wiley Online Library}
}

@book{mcdonald2013test,
  title={Test theory: A unified treatment},
  author={McDonald, Roderick P},
  year={2013},
  publisher={psychology press}
}

@article{cronbach1951coefficient,
  title={Coefficient alpha and the internal structure of tests},
  author={Cronbach, Lee J},
  journal={psychometrika},
  volume={16},
  number={3},
  pages={297--334},
  year={1951},
  publisher={Springer-Verlag}
}

@article{brauner2026charting,
  title={Charting the AI perception gap: divergent views on risk, benefit, and value between experts and the public challenge the societal acceptance of AI},
  author={Brauner, Philipp and Glawe, Felix and Liehner, Gian Luca and Vervier, Luisa and Ziefle, Martina},
  journal={AI \& society},
  volume={41},
  pages={8125--8152},
  year={2026},
  publisher={Springer}
}

@inproceedings{birhane2022power,
  title={Power to the people? Opportunities and challenges for participatory AI},
  author={Birhane, Abeba and Isaac, William and Prabhakaran, Vinodkumar and Diaz, Mark and Elish, Madeleine Clare and Gabriel, Iason and Mohamed, Shakir},
  booktitle={Proceedings of the 2nd ACM Conference on Equity and Access in Algorithms, Mechanisms, and Optimization},
  pages={1--8},
  year={2022}
}

@inproceedings{nass1994computers,
  title={Computers are social actors},
  author={Nass, Clifford and Steuer, Jonathan and Tauber, Ellen R},
  booktitle={Proceedings of the SIGCHI conference on Human factors in computing systems},
  pages={72--78},
  year={1994}
}

@inproceedings{mun2024particip,
  title={Particip-ai: A democratic surveying framework for anticipating future ai use cases, harms and benefits},
  author={Mun, Jimin and Jiang, Liwei and Liang, Jenny and Cheong, Inyoung and DeCario, Nicole and Choi, Yejin and Kohno, Tadayoshi and Sap, Maarten},
  booktitle={Proceedings of the AAAI/ACM Conference on AI, Ethics, and Society},
  volume={7},
  number={1},
  pages={997--1010},
  year={2024}
}

@ARTICLE{8226850,
  author={Taylor, Sara and Jaques, Natasha and Nosakhare, Ehimwenma and Sano, Akane and Picard, Rosalind},
  journal={IEEE Transactions on Affective Computing}, 
  title={Personalized Multitask Learning for Predicting Tomorrow's Mood, Stress, and Health}, 
  year={2020},
  volume={11},
  number={2},
  pages={200-213},
  doi={10.1109/TAFFC.2017.2784832}}

@inproceedings{kian2025engagement,
  title={Engagement and Disclosures in LLM-Powered Cognitive Behavioral Therapy Exercises: A Factorial Design Comparing the Influence of a Robot vs. Chatbot Over Time},
  author={Kian, Mina and Zong, Mingyu and Fischer, Katrin and Velentza, Anna-Maria and Singh, Abhyuday and Shrestha, Kaleen and Sang, Pau and Upadhyay, Shriya and Browning, Wallace and Faruki, Misha Arif and others},
  booktitle={2025 34th IEEE International Conference on Robot and Human Interactive Communication (RO-MAN)},
  pages={1173--1180},
  year={2025},
  organization={IEEE}
}

@inproceedings{park2019model,
  title={A model-free affective reinforcement learning approach to personalization of an autonomous social robot companion for early literacy education},
  author={Park, Hae Won and Grover, Ishaan and Spaulding, Samuel and Gomez, Louis and Breazeal, Cynthia},
  booktitle={Proceedings of the AAAI conference on artificial intelligence},
  volume={33},
  number={01},
  pages={687--694},
  year={2019}
}

@article{sharma2023human,
  title={Human--AI collaboration enables more empathic conversations in text-based peer-to-peer mental health support},
  author={Sharma, Ashish and Lin, Inna W and Miner, Adam S and Atkins, David C and Althoff, Tim},
  journal={Nature Machine Intelligence},
  volume={5},
  number={1},
  pages={46--57},
  year={2023},
  publisher={Nature Publishing Group UK London}
}

@article{lee2024towards,
  title={Towards social ai: A survey on understanding social interactions},
  author={Lee, Sangmin and Li, Minzhi and Lai, Bolin and Jia, Wenqi and Ryan, Fiona and Cao, Xu and Kara, Ozgur and Boote, Bikram and Shi, Weiyan and Yang, Diyi and others},
  journal={arXiv preprint arXiv:2409.15316},
  year={2024}
}


\newpage
\section*{Appendix}

\begin{table}[h]
\centering
\small
\setlength{\tabcolsep}{4pt}
\renewcommand{\arraystretch}{1.08}
\begin{tabular}{p{0.92\linewidth}}
\hline
\textbf{AI Literacy Scale item} \\
\hline
I can identify the AI technology employed in the applications and products I use. \\
I can skillfully use AI applications or products to help me with my daily work. \\
I can choose the most appropriate AI application or product from a variety for a particular task. \\
I always comply with ethical principles when using AI applications or products. \\
I am always alert to privacy and information security issues when using AI applications or products. \\
I am always alert to the abuse of AI technology. \\
I am familiar with limitations and shortcomings of AI agents. \\

\hline
\end{tabular}
\caption{AI Literacy Scale items.}
\label{tab:ai_literacy_scale}
\end{table}

\begin{figure}[h]
    \centering
    \fbox{\includegraphics[width=0.5\linewidth]{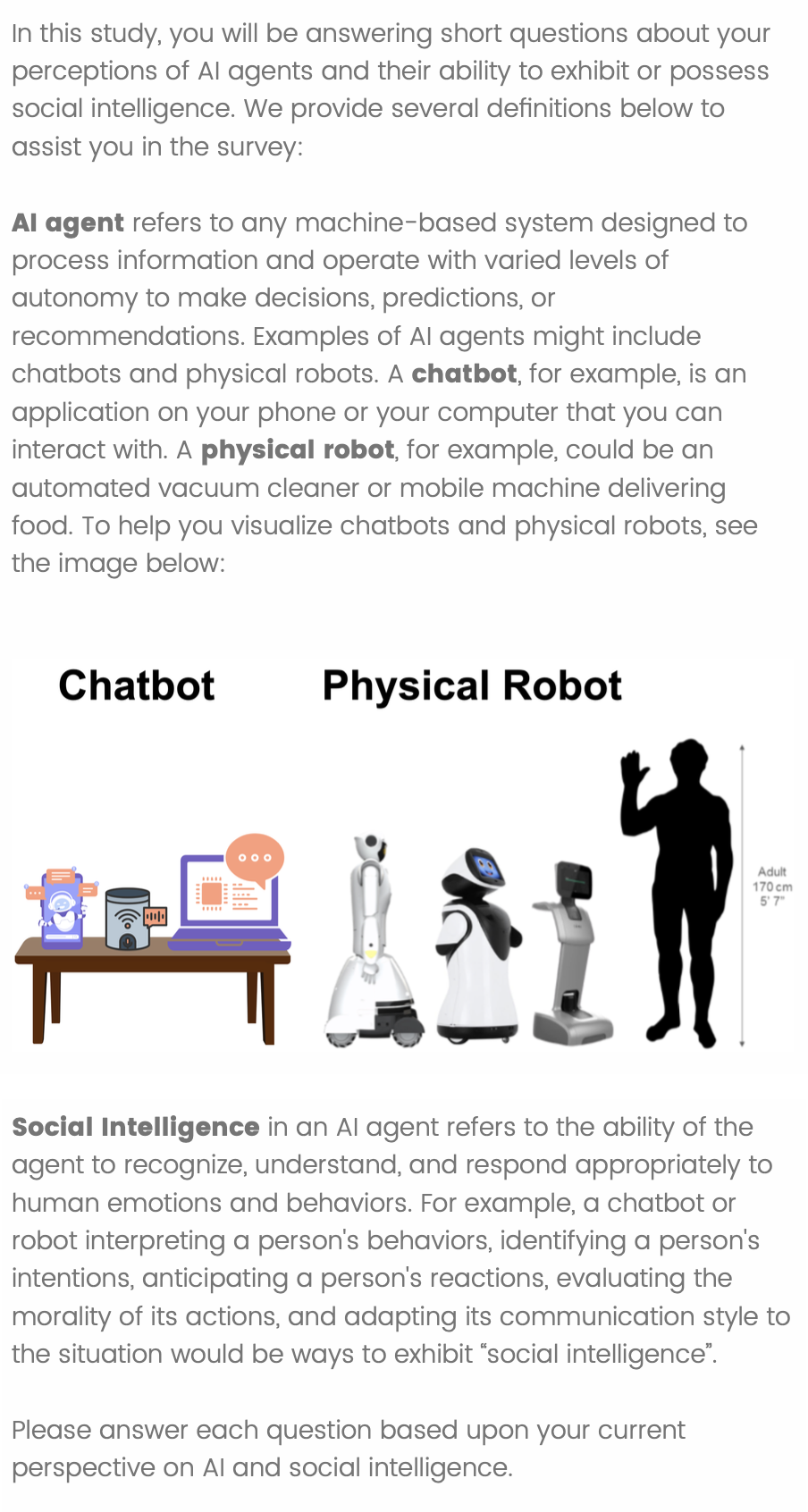}}
    \caption{Definition screen shown before the survey questions. The screen provided definitions of \textit{AI agent} and \textit{social intelligence} and showed visual anchors for chatbots and physical robots. Robot image anchors were selected from the MUFaSAA dataset \cite{dennler2023design}.}
    \label{fig:anchors}
\end{figure}

\begin{figure}[h]
    \centering
    \fbox{\includegraphics[width=0.5\linewidth]{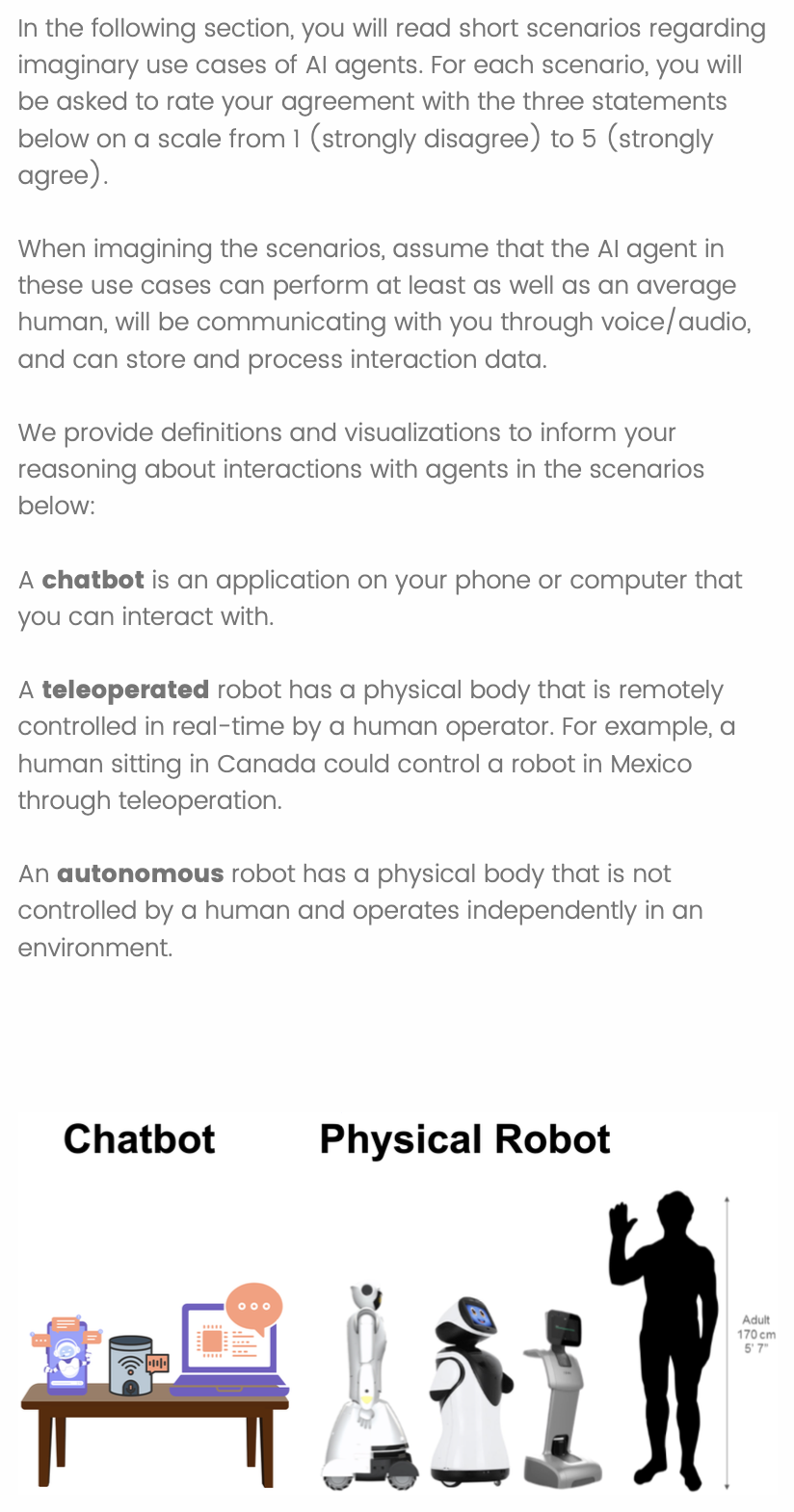}}
    \caption{Instructions, definitions, and visual examples presented to participants before providing acceptability ratings in Block C. Chatbot and robot image anchors provided before Block C are identical to the images provided at the beginning of the survey.}
    \label{fig:block_4}
\end{figure}

\begin{table*}[t]
\centering
\footnotesize
\setlength{\tabcolsep}{4pt}
\renewcommand{\arraystretch}{1.08}
\begin{tabular}{p{0.42\linewidth}rrrrrr}
\hline
\textbf{Item label} & \textbf{M} & \textbf{Str. dis.} & \textbf{Dis.} & \textbf{Neutral} & \textbf{Agree} & \textbf{Str. agree} \\
\hline
Are pleasant and sociable to interact with & 3.95 & 1.5 & 3.5 & 15.1 & 58.3 & 21.6 \\
Recognize and respond appropriately to human emotion & 3.66 & 3.0 & 9.0 & 25.5 & 44.5 & 18.0 \\
Have social intelligence & 3.57 & 8.0 & 6.5 & 22.0 & 48.0 & 15.5 \\
Take initiative and offer useful assistance & 3.55 & 6.0 & 13.0 & 18.0 & 46.0 & 17.0 \\
Build rapport and trust with humans & 3.57 & 5.0 & 8.5 & 26.0 & 46.0 & 14.5 \\
Feel natural and human-like & 3.46 & 5.5 & 14.5 & 22.5 & 43.5 & 14.0 \\
Understand what is ethically right and wrong & 3.39 & 8.5 & 11.5 & 29.5 & 33.5 & 17.0 \\
Act with purpose and follow their own intentions & 2.94 & 10.6 & 24.1 & 31.7 & 28.1 & 5.5 \\
\hline
\end{tabular}
\caption{Item-level summary statistics for the Perceived AI Social Intelligence Measure. Participants were asked to rate statements about the extent to which current AI agents have each ability. M is the mean Likert rating for the participant group. Likert columns report percentages of participants who selected each Likert response option: Str. dis. = Strongly disagree; Dis. = Disagree; Str. agree = Strongly agree.}
\label{tab:rq1_item_summary}
\end{table*}

\begin{table*}[t]
\centering
\small
\begin{tabular}{p{0.60\linewidth}rrr}
\hline
Item correlated with ``Have social intelligence'' & N & $\rho$ & FDR $p$ \\
\hline
Recognize and respond appropriately to human emotion & 200 & 0.636 & $<.001$ \\
Feel natural and human-like & 200 & 0.560 & $<.001$ \\
Are pleasant and sociable to interact with & 199 & 0.555 & $<.001$ \\
Understand what is ethically right and wrong & 200 & 0.549 & $<.001$ \\
Build rapport and trust with humans & 200 & 0.530 & $<.001$ \\
Take initiative and offer useful assistance & 200 & 0.467 & $<.001$ \\
Act with purpose and follow their own intentions & 199 & 0.367 & $<.001$ \\
\hline
\end{tabular}
\caption{Spearman correlations between the holistic social intelligence item and the seven related agent capability items in survey Block B. N is the available-case count for each item pair (one participant chose not to respond to the ``Are pleasant and sociable to interact with'' and ``Act with purpose and follow their own intentions'' items). All seven correlations remained significant after Benjamini-Hochberg false-discovery-rate (FDR) correction \cite{benjamini1995controlling} across these seven comparisons.}
\label{tab:rq1_item_correlations}
\end{table*}

\begin{table*}[t]
\centering
\scriptsize
\setlength{\tabcolsep}{0pt}
\renewcommand{\arraystretch}{1.05}
\begin{tabular}{p{0.10\linewidth}p{0.10\linewidth}p{0.80\linewidth}}
\hline
\textbf{Scenario label} & \textbf{Agent type} & \textbf{Participant-facing scenario} \\
\hline
Work support & Chatbot & You have a chatbot on a home device that notices when you are overwhelmed with work and suggests breaks. \\
Fall alert & Chatbot & You have a chatbot on a home device for an elderly parent that alerts you when they fall and appear to need assistance. \\
Wayfinding & Chatbot & A check-in kiosk at a hospital has a chatbot that notices when you appear confused and provides you with direction. \\
Distress alert & Chatbot & As a hospital patient, you interact with a chatbot that monitors your emotional state to alert your care team about any severe distress. \\
Work support & Auton. robot & You have an autonomous robot at home that notices when you are overwhelmed with work and suggests breaks. \\
Fall alert & Auton. robot & You have an autonomous robot at home for an elderly parent that alerts you when they fall and appear to need assistance. \\
Wayfinding & Auton. robot & A hospital has an autonomous robot that notices when you appear confused and approaches you to provide direction. \\
Distress alert & Auton. robot & As a hospital patient, you interact with an autonomous robot that monitors your emotional state to alert your care team about any severe distress. \\
Work support & Teleop. robot & You have a teleoperated robot at home that notices when you are overwhelmed with work and suggests breaks. \\
Fall alert & Teleop. robot & You have a teleoperated robot at home for an elderly parent that alerts you when they fall and appear to need assistance. \\
Wayfinding & Teleop. robot & A hospital has a teleoperated robot that notices when you appear confused and approaches you to provide direction. \\
Distress alert & Teleop. robot & As a hospital patient, you interact with a teleoperated robot that monitors your emotional state to alert your care team about any severe distress. \\
\hline
\end{tabular}
\caption{Wording for the 12 Social-AI acceptability scenario conditions. Wording was minimally adapted for each agent type to ensure plausibility. Each scenario was followed by the same three rating statements: whether participants would actively seek out the service for themselves, actively avoid the service for themselves, and be glad the service existed for others who wanted to use it.}
\label{tab:scenario_wording_appendix}
\end{table*}

\begin{table*}[t]
\centering
\footnotesize
\setlength{\tabcolsep}{3pt}
\renewcommand{\arraystretch}{1.12}
\begin{tabularx}{\linewidth}{@{}>{\raggedright\arraybackslash}p{.19\linewidth}>{\raggedright\arraybackslash}p{.15\linewidth}XXX@{}}
\toprule
Scenario & Agent type & Seek-for-self M [95\% CI] & Avoid-for-self M [95\% CI] & Glad-for-others M [95\% CI] \\
\midrule
Home: fall alert & Chatbot & 0.71 [0.67, 0.75] & 0.27 [0.23, 0.31] & 0.86 [0.82, 0.89] \\
Home: fall alert & Autonomous robot & 0.67 [0.63, 0.72] & 0.29 [0.25, 0.33] & 0.83 [0.80, 0.86] \\
Home: fall alert & Teleoperated robot & 0.63 [0.59, 0.67] & 0.34 [0.29, 0.38] & 0.84 [0.81, 0.87] \\
Hospital: wayfinding & Chatbot & 0.63 [0.59, 0.68] & 0.30 [0.26, 0.34] & 0.79 [0.76, 0.83] \\
Hospital: wayfinding & Autonomous robot & 0.57 [0.52, 0.61] & 0.37 [0.33, 0.41] & 0.77 [0.74, 0.81] \\
Hospital: wayfinding & Teleoperated robot & 0.53 [0.49, 0.58] & 0.35 [0.31, 0.40] & 0.75 [0.71, 0.78] \\
Hospital: distress alert & Chatbot & 0.55 [0.51, 0.60] & 0.40 [0.35, 0.45] & 0.75 [0.72, 0.79] \\
Hospital: distress alert & Autonomous robot & 0.53 [0.49, 0.58] & 0.39 [0.35, 0.44] & 0.75 [0.71, 0.78] \\
Hospital: distress alert & Teleoperated robot & 0.53 [0.49, 0.57] & 0.39 [0.35, 0.44] & 0.75 [0.72, 0.79] \\
Home: work support & Chatbot & 0.53 [0.49, 0.58] & 0.41 [0.36, 0.45] & 0.74 [0.70, 0.78] \\
Home: work support & Autonomous robot & 0.49 [0.44, 0.53] & 0.43 [0.38, 0.48] & 0.72 [0.68, 0.76] \\
Home: work support & Teleoperated robot & 0.43 [0.39, 0.48] & 0.48 [0.43, 0.53] & 0.67 [0.62, 0.71] \\
\bottomrule
\end{tabularx}
\caption{Means and 95\% confidence intervals for the 12 scenario conditions, on the normalized 0--1 scale. Higher seek and glad ratings indicate more favorable judgments; higher avoidance indicates greater active avoidance. Seek and glad use N = 200 in every condition. Avoidance uses N = 199 except for teleoperated hospital distress alert (N = 200).}
\label{tab:rq3_condition_means}
\end{table*}

\begin{table*}[t]
\centering
\footnotesize
\setlength{\tabcolsep}{3pt}
\renewcommand{\arraystretch}{1.12}
\begin{tabularx}{\linewidth}{@{}>{\raggedright\arraybackslash}X>{\raggedright\arraybackslash}p{.20\linewidth}rrrr@{}}
\toprule
Scenario & Agent type & N & Mean gap & W & Holm $p$ \\
\midrule
Home: fall alert & Chatbot & 200 & 0.14 & 341.0 & $<.001$ \\
Home: fall alert & Autonomous robot & 200 & 0.16 & 371.5 & $<.001$ \\
Home: fall alert & Teleoperated robot & 200 & 0.21 & 150.0 & $<.001$ \\
Hospital: wayfinding & Chatbot & 200 & 0.16 & 378.0 & $<.001$ \\
Hospital: wayfinding & Autonomous robot & 200 & 0.21 & 385.0 & $<.001$ \\
Hospital: wayfinding & Teleoperated robot & 200 & 0.22 & 34.5 & $<.001$ \\
Hospital: distress alert & Chatbot & 200 & 0.20 & 319.5 & $<.001$ \\
Hospital: distress alert & Autonomous robot & 200 & 0.21 & 319.0 & $<.001$ \\
Hospital: distress alert & Teleoperated robot & 200 & 0.22 & 225.0 & $<.001$ \\
Home: work support & Chatbot & 200 & 0.21 & 433.5 & $<.001$ \\
Home: work support & Autonomous robot & 200 & 0.23 & 432.0 & $<.001$ \\
Home: work support & Teleoperated robot & 200 & 0.23 & 468.0 & $<.001$ \\
Overall (descriptive) & All types & 200 & 0.20 & -- & -- \\
\bottomrule
\end{tabularx}
\caption{The Support-Adoption gap is computed as Glad-for-others minus Seek-for-self. Each condition has 200 paired observations. Two-sided Wilcoxon signed-rank tests omit zero differences; Holm correction covers 12 tests. The Overall descriptive mean summarizes 2,400 paired scenario observations from 200 participants. Condition means range from 0.14 to 0.23 after rounding.}
\label{tab:rq3_gap}
\end{table*}

\begin{table*}[t]
\centering
\footnotesize
\setlength{\tabcolsep}{4pt}
\renewcommand{\arraystretch}{1.12}
\begin{tabularx}{\linewidth}{@{}lXrlrr@{}}
\toprule
Outcome & Model & Estimate & 95\% CI & Raw $p$ & BH $p$ \\
\midrule
Seek-for-self & Unadjusted & 0.146 & [0.098, 0.195] & $<.001$ & $<.001$ \\
Seek-for-self & Adjusted for prior use & 0.103 & [0.049, 0.157] & $<.001$ & $<.001$ \\
Avoid-for-self & Unadjusted & -0.117 & [-0.170, -0.064] & $<.001$ & $<.001$ \\
Avoid-for-self & Adjusted for prior use & -0.072 & [-0.129, -0.016] & $0.013$ & $0.013$ \\
Glad-for-others & Unadjusted & 0.120 & [0.068, 0.171] & $<.001$ & $<.001$ \\
Glad-for-others & Adjusted for prior use & 0.085 & [0.039, 0.131] & $<.001$ & $<.001$ \\
\bottomrule
\end{tabularx}
\caption{Associations between self-reported AI literacy and mean acceptability ratings, before and after accounting for chatbot and robot use frequency. Estimates represent differences in acceptability ratings on the 0-1 scale associated with a one-point increase in AI literacy on its original 1-5 scale. Adjusted models include chatbot and robot use frequency as categorical variables. Linear regression models use HC3 robust standard errors and t-based 95\% CIs. BH $p$ denotes Benjamini-Hochberg-adjusted p-values.}
\label{tab:literacy_prior_use}
\end{table*}

\begin{table*}[t]
\centering
\footnotesize
\setlength{\tabcolsep}{3pt}
\renewcommand{\arraystretch}{1.12}
\begin{tabularx}{\linewidth}{@{}>{\raggedright\arraybackslash}Xlrrr@{}}
\toprule
Participant measure & Mean acceptability rating & N & $\rho$ & BH $p$ \\
\midrule
Self-reported AI literacy & Seek-for-self & 200 & 0.369 & $<.001$ \\
Self-reported AI literacy & Avoid-for-self & 200 & -0.293 & $<.001$ \\
Self-reported AI literacy & Glad-for-others & 200 & 0.388 & $<.001$ \\
Perceived social intelligence & Seek-for-self & 200 & 0.404 & $<.001$ \\
Perceived social intelligence & Avoid-for-self & 200 & -0.348 & $<.001$ \\
Perceived social intelligence & Glad-for-others & 200 & 0.367 & $<.001$ \\
\bottomrule
\end{tabularx}
\caption{Associations of self-reported AI literacy and perceived social intelligence with participants' mean acceptability ratings across scenarios. Positive correlations indicate higher ratings as the participant measure increases. Perceived social intelligence is the agreement rating with the statement ``Current AI agents have social intelligence.'' Two-sided $p$-values are adjusted using Benjamini-Hochberg correction.}
\label{tab:rq3_secondary_correlations}
\end{table*}

\begin{table*}[t]
\centering
\footnotesize
\setlength{\tabcolsep}{3pt}
\renewcommand{\arraystretch}{1.12}
\begin{tabularx}{\linewidth}{@{}>{\raggedright\arraybackslash}p{.24\linewidth}rXXX@{}}
\toprule
Group & N & Seek-for-self M [95\% CI] & Avoid-for-self M [95\% CI] & Glad-for-others M [95\% CI] \\
\midrule
\multicolumn{5}{l}{\textit{AI literacy}} \\
Q1 lowest & 53 & 0.45 [0.39, 0.51] & 0.44 [0.38, 0.50] & 0.69 [0.64, 0.74] \\
Q2 & 49 & 0.54 [0.47, 0.61] & 0.41 [0.34, 0.48] & 0.72 [0.66, 0.79] \\
Q3 & 57 & 0.59 [0.54, 0.65] & 0.36 [0.30, 0.42] & 0.79 [0.75, 0.83] \\
Q4 highest & 41 & 0.72 [0.65, 0.79] & 0.24 [0.18, 0.30] & 0.89 [0.85, 0.93] \\
\midrule
\multicolumn{5}{l}{\textit{Holistic social intelligence}} \\
Strongly disagree & 16 & 0.29 [0.22, 0.37] & 0.60 [0.48, 0.72] & 0.55 [0.43, 0.67] \\
Disagree & 13 & 0.53 [0.38, 0.67] & 0.38 [0.23, 0.54] & 0.75 [0.66, 0.85] \\
Neutral & 44 & 0.51 [0.44, 0.58] & 0.41 [0.34, 0.48] & 0.71 [0.65, 0.77] \\
Agree & 96 & 0.59 [0.54, 0.63] & 0.36 [0.32, 0.40] & 0.79 [0.76, 0.83] \\
Strongly agree & 31 & 0.75 [0.67, 0.83] & 0.20 [0.13, 0.28] & 0.88 [0.83, 0.93] \\
\bottomrule
\end{tabularx}
\caption{Descriptive means and confidence intervals for Figure~\ref{fig:ai_literacy_acceptability}. Participants are grouped by AI literacy quartile (top) and grouped by agreement with the statement ``Current AI agents have social intelligence'' (below). Equal literacy scores are kept in the same quartile, resulting in unequal group sizes.}
\label{tab:rq3_predictor_group_means}
\end{table*}

\begin{table*}[t]
\centering
\footnotesize
\setlength{\tabcolsep}{3pt}
\renewcommand{\arraystretch}{1.12}
\begin{tabularx}{\linewidth}{@{}l>{\raggedright\arraybackslash}Xrrrr@{}}
\toprule
Outcome & Effect & df & $\chi^2$ & $p$ & Ratings \\
\midrule
Seek & scenario & 3 & 265.15 & $<.001$ & 2400 \\
Seek & agent type & 2 & 56.76 & $<.001$ & 2400 \\
Seek & scenario $\times$ agent type & 6 & 9.86 & $.131$ & 2400 \\
Avoid & scenario & 3 & 150.56 & $<.001$ & 2389 \\
Avoid & agent type & 2 & 20.25 & $<.001$ & 2389 \\
Avoid & scenario $\times$ agent type & 6 & 15.64 & $.016$ & 2389 \\
Glad & scenario & 3 & 208.90 & $<.001$ & 2400 \\
Glad & agent type & 2 & 17.25 & $<.001$ & 2400 \\
Glad & scenario $\times$ agent type & 6 & 15.01 & $.020$ & 2400 \\
\bottomrule
\end{tabularx}
\caption{Omnibus Wald tests from three separate linear mixed-effects models, each including scenario context, agent type, their interaction, and a participant random intercept. Scenario and agent-type tests average across levels of the other factor. The nine p-values are unadjusted. All models converged without warnings.}
\label{tab:rq3_omnibus}
\end{table*}

\begin{table*}[t]
\centering
\footnotesize
\setlength{\tabcolsep}{3pt}
\renewcommand{\arraystretch}{1.12}
\begin{tabularx}{\linewidth}{@{}>{\raggedright\arraybackslash}Xrrrrr@{}}
\toprule
Condition & Speech cont. & Speech active & Video cont. & Video active & Blank N \\
\midrule
Autonomous/local & 23.0 & 76.5 & 19.0 & 63.0 & 31 \\
Autonomous/cloud & 9.0 & 68.5 & 6.0 & 46.0 & 57 \\
Teleoperated/local & 13.5 & 66.0 & 10.0 & 52.0 & 53 \\
Teleoperated/cloud & 6.5 & 61.0 & 4.5 & 42.0 & 72 \\
\bottomrule
\end{tabularx}
\caption{Recorded selections for data a Social-AI robot in the home could use and store. Selection columns are percentages out of all 200 participants; multiple selections were permitted. Active means only during active interaction; cont. means continuous sensing. Blank N counts completely blank responses.}
\label{tab:privacy_preferences_appendix}
\end{table*}

\begin{table*}[t]
\centering
\footnotesize
\setlength{\tabcolsep}{3pt}
\renewcommand{\arraystretch}{1.12}
\begin{tabularx}{\linewidth}{@{}>{\raggedright\arraybackslash}Xlrrrr@{}}
\toprule
Condition & Modality & N & Active only & Cont. only & Holm $p$ \\
\midrule
Autonomous/local & speech & 200 & 120 & 13 & $<.001$ \\
Autonomous/local & video & 200 & 100 & 12 & $<.001$ \\
Autonomous/cloud & speech & 200 & 123 & 4 & $<.001$ \\
Autonomous/cloud & video & 200 & 85 & 5 & $<.001$ \\
Teleoperated/local & speech & 200 & 115 & 10 & $<.001$ \\
Teleoperated/local & video & 200 & 92 & 8 & $<.001$ \\
Teleoperated/cloud & speech & 200 & 111 & 2 & $<.001$ \\
Teleoperated/cloud & video & 200 & 79 & 4 & $<.001$ \\
\bottomrule
\end{tabularx}
\caption{Two-sided exact McNemar comparisons of active-interaction versus continuous-sensing selections. Active only and cont. only are the discordant participant counts. Holm correction covers eight comparisons (speech and video in four robot/storage conditions). Every adjusted p-value is below .001.}
\label{tab:privacy_paired_tests}
\end{table*}

\begin{table*}[t]
\centering
\scriptsize
\setlength{\tabcolsep}{3pt}
\renewcommand{\arraystretch}{1.10}
\begin{tabular}{p{0.62\linewidth}p{0.30\linewidth}}
\hline
\textbf{Participant Response} & \textbf{Assigned RQ2 code(s)} \\
\hline
There have been times when I have responded sarcastically and it will understand that I don't literally mean what I'm saying. I don't remember the specific context, but the chat bot had explained something negative in a matter-of-fact way, and when I thanked it sarcastically ``Wow. Thanks for that.'' it picked up my meaning and apologized for the sterile tone in the prior message & current context awareness; navigating conversations; social norms \\
\hline
It listened to what I was saying and didn't jump to conclusions. When I got mad it didn't try to flatter me or overdo empathy, but it addressed my concerns in a direct way. It also kept the conversation on track. It answered my questions clearly and it didn't waste my time. I think it kept it professional but still human like. It didn't get offended when i snapped at it. & humanlike communication; logical structure; navigating conversations; recognizing emotions \\
\hline
The chatbot seemed to pick up on pieces of the puzzle that I did not tell it, and gave me suggestions that I hadn't thought nor the people that I confided to in first. It felt like I was talking to a human. & building trust; current context awareness; helpful; humanlike communication \\
\hline
It knows my habits, preferences, order history, news interests, favorite music, and a lot of things that make it feel socially alert to my needs. I like that it has a lot more capability than previous Alexa versions, it can really do a lot and also feels like maybe it can be too intelligent at times (it has surprised me at times with its comments, similar to some human interactions that raise emotion). When the agent can accurately describe my previous history and also my current surroundings in detail, it makes me feel it has a lot of intelligence. It also gives me regular schedule reminders and can update things immediately if I need them to, which makes it feel in-tune with all my daily tasks. It sometimes remembers things even better than I can, its intelligence can be humbling at times. & helpful; humanlike communication; longitudinal awareness \\
\hline
I can't remember exactly what Gemini said, but it started out with a congratulations and then went on to reflect back to me how I was feeling. It then went on to ask me questions related to the job in a way that felt like it not only was proud of me and excited I got the job, but also interested in how I was feeling. & expressing emotions; expressing empathy; navigating conversations; recognizing emotions \\
\hline
The chat bot listed three short steps like how to submit a dispute, what to write in the dispute and when to expect a refund after submission. The chat bot used simple pauses between items, and added an approximation on when I should hear back from the third party, which I didn't ask for. & helpful; logical structure; task competence \\
\hline
No ability can convince me that an AI has social intelligence. A robot cannot feel things and if a person is feeling any way the AI would not be able to help because of no empathy, sympathy, or understanding. & rejecting the premise \\
\hline
\end{tabular}
\caption{Representative RQ2 responses and assigned codes. Responses could receive multiple codes.}
\label{tab:rq2_representative_quotes_appendix}
\end{table*}

\begin{table*}[t]
\centering
\scriptsize
\setlength{\tabcolsep}{3pt}
\renewcommand{\arraystretch}{1.10}
\begin{tabular}{p{0.62\linewidth}p{0.30\linewidth}}
\hline
\textbf{Participant Response} & \textbf{Assigned RQ4 code(s)} \\
\hline
I have no concerns or reservations about socially-intelligent AI. I like that they exist and seem to be getting better all of the time. I look forward to a time when they can help even more. & no concerns \\
\hline
I have a few concerns about socially intelligent AI. I think the biggest is privacy, where systems that monitor emotions or behavior (collecting highly sensitive data) and it's not always clear who controls it or how it's used. I think there's also a risk of overreliance, where people start learning on AI, or leaning on AI for emotional support or decisions instead of others. Finally, I think the systems can misread situations, and in places like healthcare, a wrong interpretation could often lead to the wrong response. & data privacy; incorrect responses; overreliance; reduce human connection \\
\hline
I am concerned that AI will be manipulated by bad actors in order to spread false information or mislead humans in order to enrich or empower a select group of people. & data misuse; manipulation \\
\hline
There are no guardrails in place, ethically, to use AI as wide spread as it already is, let alone being heavily involved in healthcare. It is a program built intentionally on randomness and can have legitimate uses, but right now it isn't dependable for serious matters, and is basically just a tool to avoid a need for human labor. & accountability governance; incorrect responses; replace human jobs \\
\hline
My basic concern is that so many of the conditions or emotions which you've highlighted need to be states or conditions which we should be well aware of on our own, and yes, that requires practice and perhaps initially some counseling from another (trained) soul. Bypassing a person's ability to care for themselves - and by association others - could ironically magnify and increase social isolation for the human population. & reduce human ability; reduce human connection \\
\hline
I have concerns when it comes to trusting AI agents with health and stress detection. I do not feel comfortable enough with AI agents to accurately detect human emotions. & overestimate functional ability; overestimate social ability \\
\hline
I think human controlled robots inside homes are a bit of scary thing. I wouldn't really consent to that kind of access to my home. I think this applies to autonomous AI too in a lot of ways. I'd have to think on it more if i'd be more accepting of it. & data privacy; embodied discomfort \\
\hline
\end{tabular}
\caption{Representative RQ4 responses and assigned codes. Responses could receive multiple codes.}
\label{tab:rq4_representative_quotes_appendix}
\end{table*}

\end{document}